\documentclass[twocolumn]{aastex63}

\UseRawInputEncoding
\usepackage{graphicx}
\usepackage{amsmath} 
\usepackage{natbib, twoopt}

\submitjournal{ApJ}

\begin{document}


\title{The star formation history of Eridanus II: on the role of SNe feedback in the quenching of ultra-faint dwarf galaxies.} 

\footnote{Based on observations made with the NASA/ESA Hubble Space Telescope, obtained at the Space Telescope Science Institute, which is operated by the Association of Universities for Research in Astronomy, Inc., under NASA contract NAS 5-26555. These observations are associated with programs 14224 and 14234.}

\footnote{Accepted December 29, 2020}

\correspondingauthor{Carme Gallart}
\email{carme.gallart@iac.es}

\author[0000-0001-6728-806X]{C. Gallart}
\affiliation{Instituto de Astrof\'\i sica de Canarias, 38205 La Laguna, Tenerife, Spain}
\affiliation{Departamento de Astrof\'isica, Universidad de La Laguna, E-38200 La Laguna, Tenerife, Spain}
\author{M. Monelli}
\affiliation{Instituto de Astrof\'\i sica de Canarias, 38205 La Laguna, Tenerife, Spain}
\affiliation{Departamento de Astrof\'isica, Universidad de La Laguna, E-38200 La Laguna, Tenerife, Spain}
\author[0000-0001-6984-4795]{T. Ruiz-Lara}
\affiliation{Kapteyn Astronomical Institute, University of Groningen, Landleven 12, 9747 AD Groningen, The Netherlands}
\author[0000-0002-0882-7702]{A. Calamida}
\affiliation{Space Telescope Science Institute, 3700 San Martin Drive, Baltimore, MD 21218}
\author{S. Cassisi} 
\affiliation{INAF - Osservatorio Astronomico d'Abruzzo, Via M. Maggini, s/n, I-64100, Teramo, Italy}
\affiliation{INFN, Largo B. Pontecorvo 3, 56127, Pisa, Italy} 
\author{M. Cignoni}
\affiliation{INFN, Largo B. Pontecorvo 3, 56127, Pisa, Italy} 
\affiliation{Department of Physics - University of Pisa, Largo Pontecorvo, 3 Pisa, 56127, Italy } 
\affiliation{INAF-Osservatorio di Astrofisica e Scienza dello Spazio, Via Gobetti 93/3, 40129, Bologna, Italy}
\author{J. Anderson}
\affiliation{Space Telescope Science Institute, 3700 San Martin Drive, Baltimore, MD 21218}
\author{G. Battaglia}
\affiliation{Instituto de Astrof\'\i sica de Canarias, 38205 La Laguna, Tenerife, Spain}
\affiliation{Departamento de Astrof\'isica, Universidad de La Laguna, E-38200 La Laguna, Tenerife, Spain}
\author{J. R. Bermejo-Climent}
\affiliation{INAF-Osservatorio di Astrofisica e Scienza dello Spazio, Via Gobetti 93/3, 40129, Bologna, Italy}
\affiliation{INFN, Sezione di Bologna, via Irnerio 46, I-40126 Bologna, Italy}
\affiliation{Departamento de Astrof\'isica, Universidad de La Laguna, E-38200 La Laguna, Tenerife, Spain}
\author{E. J. Bernard}
\affiliation{Instituto de Astrof\'\i sica de Canarias, 38205 La Laguna, Tenerife, Spain}
\author[0000-0002-9144-7726]{C. E. Mart\'\i nez-V\'azquez}
\affiliation{Cerro Tololo Inter-American Observatory, NSF's National Optical-Infrared Astronomy Research Laboratory, Casilla 603, La Serena,
Chile}
\author{L. Mayer}
\affiliation{Center for Theoretical Astrophysics and Cosmology, Institute for Computational Science, University of Zurich, Winterthurerstrasse 190, CH-8057 Zurich, Switzerland}
\author{S. Salvadori}
\affiliation{Dipartimento di Fisica e Astrofisica, Univerisitá degli Studi di Firenze, via G. Sansone 1,Sesto Fiorentino,Italy}
\affiliation{INAF/Osservatorio Astrofisico di Arcetri, Largo E. Fermi 5, Firenze, Italy}
\author{A. Monachesi}
\affiliation{Instituto de Investigaci\'on Multidisciplinar en Ciencia y Tecnolog\'ia, Universidad de La Serena, Ra\'ul Bitr\'an 1305, La Serena, Chile}
\affiliation{Departamento de Astronom\'ia, Universidad de La Serena, Av. Juan Cisternas 1200 Norte, La Serena, Chile}
\author{J. F. Navarro}
\affiliation{Department of Physics and Astronomy, University of Victoria, PO Box 3055, STN CSC, Victoria BC V8W 3P6, Canada}
\author{S. Shen}
\affiliation{Institute of Theoretical Astrophysics, University of Oslo, Sem S\ae lands vei 13, 0371, Oslo, Norway}
\author{F. Surot}
\affiliation{Instituto de Astrof\'\i sica de Canarias, 38205 La Laguna, Tenerife, Spain}
\affiliation{Departamento de Astrof\'isica, Universidad de La Laguna, E-38200 La Laguna, Tenerife, Spain}
\author{M. Tosi}
\affiliation{INAF-Osservatorio di Astrofisica e Scienza dello Spazio, Via Gobetti 93/3, 40129, Bologna, Italy}
\author{V. Bajaj}
\affiliation{Space Telescope Science Institute, 3700 San Martin Drive, Baltimore, MD 21218}
\author{G. S. Strinfellow}
\affiliation{Center for Astrophysics and Space Astronomy, University of Colorado, 389 UCB, Boulder, Colorado 80309-0389}




\begin{abstract}
Eridanus II (EriII) is an ultra-faint dwarf (UFD) galaxy (M$_V$=-7.1) located at a distance close to the Milky Way virial radius. Early shallow color-magnitude diagrams (CMD) indicated that it possibly hosted an intermediate-age or even young stellar population, which is unusual for a galaxy of this mass. In this paper, we present new ACS/HST CMDs reaching the oldest main sequence turnoff with excellent photometric precision, and derive a precise star formation history (SFH) for this galaxy through CMD-fitting. This SFH shows that the bulk of the stellar mass in Eri II formed in an extremely short star formation burst at the earliest possible time. The derived star formation rate profile has a width at half maximum of 500 Myr and reaches a value compatible with null star formation 13 Gyr ago. However, tests with mock stellar populations and with the CMD of the globular cluster M92 indicate that the star formation period could be shorter than 100 Myr.

From the quantitative determination of the amount of mass turned into stars in this early star formation burst ($\sim 2 \times 10^5 M_\odot$) we infer the number of SNe events and the corresponding energy injected into the interstellar medium. For reasonable estimates of the EriII virial mass and values of the coupling efficiency of the SNe energy, we conclude that EriII could be quenched by SNe feedback alone, thus casting doubts on the need to invoke cosmic reionization as the preferred explanation for the early quenching of old UFD galaxies.  


\end{abstract}


\section{Introduction} \label{sec:intro}

Dwarf galaxies are the most numerous type of galaxies in the present-day Universe. They are thought to be the first galaxies to form and the basic building-blocks of the stellar halos of larger galaxies \citep{Helmi2020ARAA}. They are thus, key astrophysical objects for understanding the most common mode of galaxy formation and how they relate to the build-up of larger structures in the Universe. In addition, their relatively simple nature makes them ideal test-beds of the physics that goes into galaxy formation and evolution models. Processes such as heating by the cosmic UV background and internal feedback from SNe are two mechanisms able to dramatically affect the formation and evolution of dwarf galaxies \citep[e.g.][]{MacLowFerrara1999, Ricotti2002, Kravtsov2004, Salvadori2008, Stinson2009, Sawala2010, Salvadori2014, Shen2014, Cashmore2017, Fitts2017, Jeon2017, RevazJablonka2018, Romano2019, Wheeler2019, Katz2020, Rey2020, Gelli2020}.

The relative impact of the above effects varies with the mass of the dwarf, with the largest systems potentially unaffected by reionization and less affected by SN feedback in terms of the capability of gas removal (while feedback might still be able to alter the density profile of the inner regions of their dark matter halos). Local effects are also at play if the galaxy is located in a dense environment or close to a massive galaxy \citep[see][for a review]{Mayer2010}: not only can the local UV flux be an order of magnitude larger than the cosmic average, but the ram-pressure stripping and the tidal interaction can profoundly alter the
properties of the gaseous, stellar and dark matter components of the dwarf galaxy. However, recent hydro-dynamical zoom-in simulations of massive high-z Lyman Break Galaxies and its dwarf companions \citep{Pallottini2017}, have shown that SN feedback is a key physical mechanism driving the evolution of dwarf galaxy satellites, even in these extremely dense and biased environments \citep{Gelli2020}.

It is expected that reionization must leave some signature on the stellar content of dwarf galaxies below a certain mass range, and particularly in their early star formation history (SFH). However, theoretical models have made contrasting predictions on what to expect. Cosmological models and simulations that include star forming minihalos, i.e. the first galaxies \citep[e.g.][]{Bromm2011}, found that owing to SN feedback the star formation can be suppressed prior to the end of reionization in small systems with dark matter mass $<10^8M_{\odot}$, which are possibly associated to present-day ultra-faint dwarf  (UFD) galaxies \citep[e.g.][]{Ricotti2005, BovillRicotti2009, SalvadoriFerrara2009, BH2015, Jeon2015}. After reionization the hot intergalactic gas ($\approx 2\times 10^4$K) cannot be accreted by these minihalos ($T_{vir} < 10^4$K), which therefore passively evolve until the present-day unless they experience major mergers \citep{Salvadori2015} or late accretion events (z $<$ 2) that can fuel new star formation activity when the intensity of the UV background decreases \citep{Babul1992, Ricotti2009}. Conversely, simulations that cannot resolve the star formation in $\rm H_2$-cooling minihalos predict that small dwarf galaxies are hosted by more massive dark matter halos, which are less prone to feedback processes and can keep on forming stars after reionization in spite of a slow decline \citep[e.g.][]{Sawala2010, Wheeler2015, Romano2019}. In conclusion, the hosting halo mass seems to be the main driver of the evolution of dwarf galaxies. For example, \citet{BenitezLlambay2015} have discussed that the varied SFHs observed in nearby dwarf galaxies may be explained by a combination of the diversity of accretion histories and the effects of cosmic reionization and feedback on the SFH, which critically depend on the mass acquired by the object at the time when reionization is complete. In turn, \citet{Gallart2015} have suggested that this may be related to the environment where a dwarf galaxy has formed, and that therefore, this is the ultimate origin of the two main types of SFHs ({\it slow} and {\it fast}) observed in Local Group dwarf galaxies. 

The precise SFHs for Local Group galaxies obtained thanks to deep observations (particularly from the {\it Advanced Camera for Surveys}, ACS on board the {\it Hubble Space Telescope}, HST) have been essential in providing observational constraints to state-of-the-art models of galaxy formation and evolution. The Local Cosmology from Isolated dwarfs (LCID) project \citep[see][and references therein,  for a discussion of global results of the project]{Gallart2015} pioneered the use of color-magnitude diagrams (CMDs) reaching the oldest main-sequence turnoffs  (oMSTO) to obtain precise, complete SFHs for isolated Local Group dwarf galaxies. The results of this project showed that, at least in the Local Group environment, reionization alone could not have halted star formation in dwarf galaxies in the stellar mass range $10^6 - 10^7 M\odot$ \citep{Monelli2010sfhcetus, Monelli2010sfhtucana, Hidalgo2011sfhlgs3, Hidalgo2013}. Since 2005, a large number of fainter dwarf galaxies have been  discovered \citep{Willman2005b, Belokurov2006,Zucker2006a}, both as Milky Way (MW) and Andromeda galaxy satellites. They were generically named UFDs and they appear to be the least luminous, least chemically evolved and most dark-matter-dominated galaxies known. Their absolute magnitudes range from $M_V = -1.5$ (Segue I) to $M_V \simeq -8$ (Leo T and CVn I). \citet{Brown2014} have used ACS \@HST CMDs reaching well below the oMSTO in six of these galaxies and concluded that they formed  $80\%$ and $100\%$ of their stars by z $\simeq$ 6 (12.8 Gyr ago) and z $\simeq$ 3 (11.6 Gyr ago), respectively. From the similar ancient populations of these galaxies, the authors find support for the hypothesis that star formation in them was suppressed by a global outside influence such as cosmic reionization.

This conclusion, however, was challenged when the Eridanus II (Eri~II hereinafter, $M_V=-7.1$) and Hydra~II ($M_V = -4.6$) dwarf galaxies were discovered \citep{Bechtol2015, Koposov2015a, Martin2015Hydra}. Their early ground-based CMDs  showed possible indications of the presence in them of an intermediate-age, or even, young population like in the case of Leo T, 
\citep{Clementini2012, Weisz2012LeoT}. What mechanisms may have allowed star formation to substantially stretch in time in these extremely small galaxies? The extended SFH in some of them may point to the intriguing possibility that reionization may not be, after all, the cause for the star formation shut down of UFD at very early times.

\begin{deluxetable}{lcc}
\tablecaption{Basic data of Eridanus II\label{basic_data}}
\tablecolumns{3}
\tablehead{
\colhead{Parameter} & \colhead{Eri II} & \colhead{ref}
}
\startdata
RA (h:m:s, J2000) & 03:44:20.1 & (1) \\
DEC  (deg:m:s, J2000) & -−43:32:01.7 & (1)\\
M$_V$ (mag) & -7.1$\pm$ 0.3 & (1) \\
(m-M)$_0$ & 22.87$\pm0.10$& (2) \\
D$_{MW}$ (Kpc) & 375 & (2) \\
r$_h$ (arcmin) & 2.31 $\pm$  0.12 & (1) \\
r$_h$ (pc) & 252 & (1,2)\\
$v_{hel}$ (km s$^{-1}$) & 75.6$\pm$1.3,2.0$^\tablenotemark{a}$ & (3)\\
$v_{GSR}$ (km s$^{-1}$) & -66.6  & (3)\\
$\mu_{\alpha}$ (mas yr$^{-1}$) & 0.159$\pm$0.292,0.053$^\tablenotemark{a}$  & (4)\\
$\mu_{\delta}$ (mas yr$^{-1}$) & 0.372$\pm$0.34,0.053$^\tablenotemark{a}$  & (4)\\
$V_{rad}$ (km s$^{-1}$) & -71$\pm$6  & (4)\\
$V_{tan}$ (km s$^{-1}$) & 612$^{+526}_{-401}$  & (4)\\
d$_{peri}$ (Kpc) & 356$^{+26}_{-45}$ $^\tablenotemark{b}$ & (4) \\
d$_{apo}$ (Kpc) & $>$500 $^\tablenotemark{b}$ & (4) \\
$\sigma_{v_{hel}}$ (km s$^{-1}$) & 6.9 $^{+1.2}_{-0.9}$ & (3)\\
M$_*$ (M$_{\odot}$) &  $1.1\times10^5$ & (5)\\
M$_{HI}$ (M$_{\odot}$) &  $< 2.8\times10^3$ & (1)\\
M$_{1\over2}$ (M$_{\odot}$) &  1.2$^{+0.4}_{-0.3}\times10^7$ & (3) \\
M/L (M$_{\odot}$/L$_{\odot}$) & 420$^{+210}_{-140}$ & (3) \\
$\mathrm{[Fe/H]}$ & -2.38 $\pm$ 0.13 & (3)\\
$\sigma_\mathrm{[Fe/H]}$ & 0.47 $^{+0.12}_{-0.09}$ & (3)\\
\enddata
\tablenotetext{a}{The numbers refer to statistical and systematic errors, respectively.}
\tablenotetext{b}{For a MW dark matter halo virial mass M=1.6$\times$10$^{12}  M_{\odot}$; very similar values are obtained for a lower mass MW halo.}
\tablecomments{ (1) \citet{Crnojevic2016}; (2) Mart\'\i nez-V\'azquez et al. in prep; 
(3) \citet{Li2017}; (4) \citet{Fritz2018}; (5) This work }
\end{deluxetable}

In the case of Eri~II, a subsequent paper by \citet{Crnojevic2016} presented a ground-based  CMD obtained with the Magellan telescope, which reached about three magnitudes deeper than the discovery CMD, down to the oMSTO (though with relative large photometric errors). It clearly excluded the presence of a young ($\simeq$ 250 Myr old) stellar component \citep[as had been suggested by][]{Koposov2015a}, but was inconclusive as to whether a few Gyr old intermediate-age population could be present in the galaxy. This work also derived Eri II structural parameters, discussed the presence of a candidate star cluster near its center \citep[which possible existence has raised  a lot of interest as a potential unique probe of the nature of dark matter  and of the dark matter density profile of Eri~II itself, see Fig.~\ref{image},][]{Amorisco2017Cluster, Contenta2018Cluster, Marsh2019Cluster, Zoutendijk2020cluster}, and showed the absence of HI gas  associated to this galaxy \citep[see also][]{Westmeier2015}. Subsequently, spectroscopy of 28 member stars was used by \citet{Li2017} to determine their velocities and metallicities, which disclosed a stellar system with low overall metallicity but considerable metallicity dispersion, high mass-to-light ratio,  and a negative velocity with respect to the Galactic standard of rest  (see Table 1) which implies that it is moving toward the Milky Way. Finally, \cite{Fritz2018} used Gaia DR2 together with the spectroscopic velocity measurements for 12 stars to derive a systemic proper motion, albeit with large errors due to its large distance and the few (faint) stars with the necessary data. The derived orbital properties would tentatively indicate that Eri~II is now close to its pericenter and possibly bound to the MW. Table \ref{basic_data} provides a summary of the basic literature data on Eri~II.

In this paper we present a detailed and precise SFH obtained by quantitatively fitting new ACS/HST CMDs reaching the oMSTO with high photometric precision with synthetic CMDs. This SFH shows that the bulk of Eri II stellar mass formed in an extremely short burst of star formation before the epoch of reionization, thus demonstrating conclusively the absence of any substantial amount of intermediate-age stellar population. 

The paper is organised as follows: Sec.~\ref{sec:observations} and 
\ref{sec:data} describe the observations, data reduction and photometry; Sec.~\ref{sec:cmd} discusses the characteristics of the observed CMD and the hints on the Eri II stellar populations that can be extracted from the comparison with theoretical isochrones; Sec.~\ref{sec:sfh_derivation} provides a detailed explanation of the SFH derivation procedures, using two independent codes and stellar evolution model sets; Sec.~\ref{sec:sfh_comment} discusses the features present in the derived SFH and some tests using mock stellar populations that help further constraining the actual intrinsic duration of Eri II old star formation event. Finally, in Sec.~\ref{discussion} we summarise our findings and discuss their implications for dwarf galaxy formation and evolution.

\section{Observations}\label{sec:observations} 

Deep photometric data for Eri~II (R.A.=03:44:21, Dec=--43:32:00, J2000) were obtained with the ACS/WFC \citep[]{Ford1998} aboard the HST under program P.ID. 14224 (Cycle 23, P.I. C. Gallart). The observations were designed to obtain a signal-to-noise ratio $\simeq$ 40 near the magnitude level of the oMSTO, at M$_{F814W} \simeq$  +2.75. Following \citet{Stetson1993ASP}, the filter choice was based on the analysis of synthetic CMDs in the ACS bands. The (F475W-F814W) pair, due to the large color baseline, is optimal for discriminating age and metallicity differences at old age, while the relatively large widths of these filters allow keeping relatively short exposure times.

Under this program, the  galaxy was observed during six orbits, organised in two visits of three orbits each. Short and long exposures were taken in order to increase the dynamical range of our CMD. In this way, we reach the oMSTO with good photometric accuracy and precision, while maintaining unsaturated the stars at the tip of the red giant branch (RGB) and any bright blue stars that might be present. The total exposure time was 7,644 s in F475W and 7,900 s 
in F814W.  Even though six epochs per filter are clearly not sufficient to obtain periods and proper light curves for short period variable stars such as RR Lyrae or Anomalous Cepheids, we did plan the observations in order to maximize variable star discovery by splitting each orbit in at least one F475W and one F814W long exposure. Parallel exposures with the WFC3/UVIS were obtained
(R.A.=03:44:18, Dec=--43:26:07, J2000). The CMD resulting from the parallel exposures does not show any feature that can be associated to Eri~II. 

Another ACS/WFC program was devoted to image the same galaxy in cycle 23. GO program 14234 (P.I. J. Simon) used 13 orbits to observe the central region of Eri~II. Ten and sixteen exposures were collected in the F606W and F814W bands, with individual exposure times ranging from 1,220 s to 1,390 s. The total integration time was 12,830 s and 20,680 s, respectively. 

\begin{figure}[h!]
\centering\includegraphics[width=0.5\textwidth]{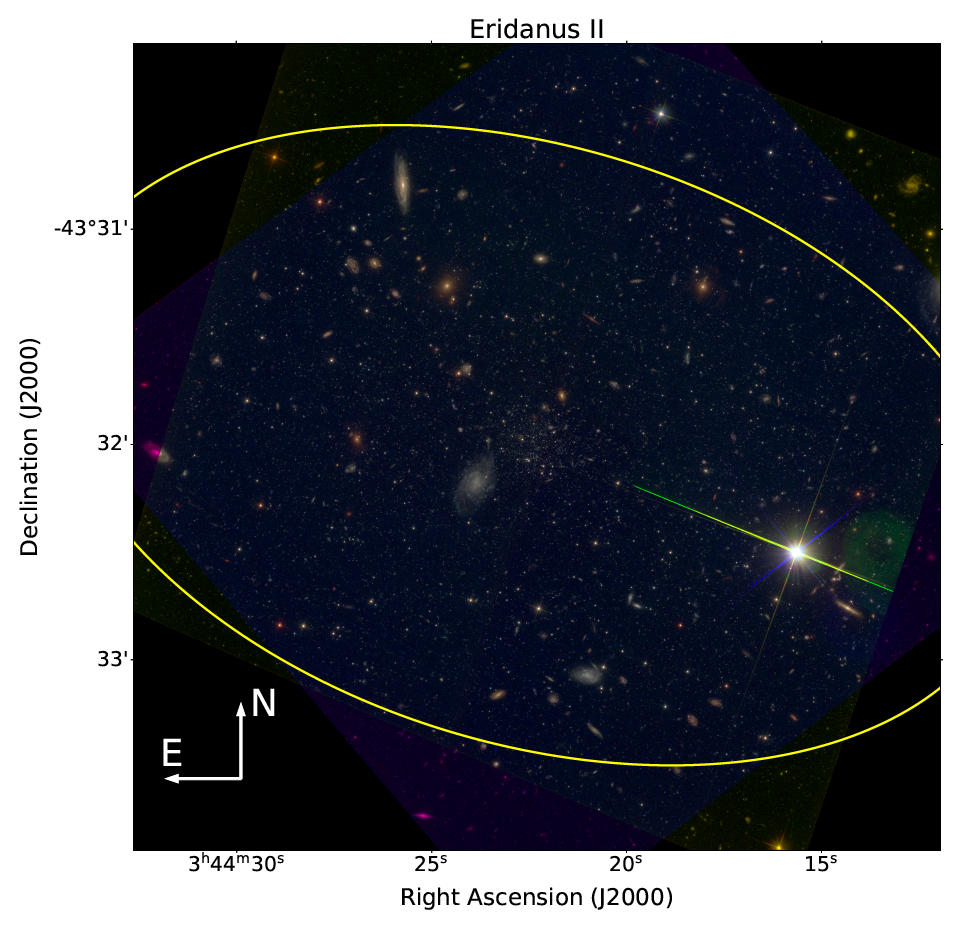}
\caption{Three color image obtained by combining the two datasets described  in the text (F475W, F606W, F814W images).  The half-light radius of Eri II  and that of the possible stellar cluster are represented by the solid- and dashed-line ellipses, respectively. NOTE: A higher resolution image will be provided for the final version of the paper.}
\label{image}
\end{figure}

Figure~\ref{image} displays a three-color image created combining the F475W, F606W and F814W images from the two datasets, which are rotated by $\simeq$ 121 degrees.  The exposures for each of the three filters were aligned to Gaia DR2 using TweakReg, and combined into corresponding mosaic images using AstroDrizzle \footnote{https://www.stsci.edu/scientific-community/software/drizzlepac }.  The three mosaics were then stacked as red, green, and blue channels to make the color image in Figure 1.  Note the presence of a large amount of background extended objects. We will use the two datasets to independently derive the SFH. A detailed analysis of the structural parameters of the galaxy and possible star cluster, together with an in-depth analysis of the variable star population of Eri II is deferred to a forthcoming paper (Mart\'\i nez-V\'azquez et al. in prep).

\section{Data reduction and photometry}\label{sec:data}

In order to track down any possible systematic issues in the SFH results due to the usage of a specific data reduction package, photometry and artificial star tests were performed both with the DAOPHOT IV/ALLFRAME suite of codes \citep{Stetson87,Stetson94} and ePSF/KS2 \citep{AndersonKing2006}, two commonly used programs to perform HST photometry. The details of the procedures followed in 
each case are described below.

\subsection{DAOPHOT IV/ALLFRAME photometry} \label{daophot}

The photometry with DAOPHOT IV/ALLFRAME was obtained following the prescriptions of \citet{Monelli2010sfhcetus}. Briefly, images are individually treated to derive optimal point spread functions (PSF) based on bright stars that are unaffected by cosmic rays and are distributed over the full field of view, in order to take into  account the spatial variations. Original \_FLC images from the ACS/HST archive pipeline  version 8.3.4 were used. The input list of sources for the simultaneous photometry performed by ALLFRAME  was generated on a stacked median image. This ensures that most, if not all, cosmic rays are filtered out and do not pollute the object lists. The background extended objects were efficiently removed using the SHARPNESS parameter ($|sha| < 0.1$).  The final selected photometric catalog was calibrated to the VEGAMAG system by using the zero points suitable for the observing epoch and available on the ACS web page. It is important to note that our catalog is calibrated to the HST photometric reference system CALSPEC v10 \footnote{https://www.stsci.edu/hst/instrumentation/reference-data-for-calibration-and-tools/astronomical-catalogs/calspec}.

Given the rotation between the two datasets, they have been independently reduced, and cross-matched a posteriori, after correcting the coordinate system for geometric distortions. The final list includes 8,350 and 12,730 bona-fide stars in common between the $F475W$ and the $F814W$ filters, and between $F606W$
and the $F814W$ filters, respectively.

In order to have an optimal sample of artificial stars to simulate the observational errors in the synthetic CMD used to derive the SFH (see Section~\ref{sec:sfh_derivation}), we adopted a synthetic CMD to create the input artificial stars list. Stars were uniformly distributed within a range of ages 2.5 $<$ t $<$ 13.5 Gyr and metallicities 0.0001 $<$ Z $<$ 0.002 to broadly cover the observed range of colours and magnitudes. Artificial stars tests were performed independently for the two data sets, simulating 10$^6$ stars per chip in $\sim$300 iterations, distributing the stars in a grid of equilateral triangles of side length equal to ($2\times PSF_{radius} +1$ px) to ensure that the photometry of artificial stars is only affected by neighbour real stars, in addition to image characteristics and defects \citep{Gallart1999data}. The stars are injected in each original images scaling for individual zero points using the corresponding PSF previously used for the data reduction. For a more realistic simulation of the photometric process, the recovery of the synthetic stars is not made with the same PSF, which somehow would match perfectly the input one and therefore would somehow underestimate the errors. Instead, during the reduction process the PSF is recalculated assuming the same list of stars of the original image but using only a randomly selected fraction of stars between 80 and 95\%. Moreover, to take into account possible uncertainties associated to the flat-field correction, we added to the input synthetic magnitudes a Gaussian error with a dispersion of 0.01 mag. This will simulate the $\approx$ 1\% flat-field error of the observations \citep{Brown2014}.

\subsection{ePSF/KS2 photometry} \label{ePSF}

Photometry is performed on the \_FLC images after correcting them for geometric distortion according to the prescription of \citet{AndersonKing2006}. Then, we carried out PSF photometry on the individual images with the routine {\it hst2xym} and by using standard PSF libraries that account for the spatial variation across the detector \citep{AndersonKing2006}. In addition, we took into account the temporal variation of the PSF by calculating a perturbation correction of the standard PSF and an individual PSF for each image. In order to increase the depth of the photometric catalog, we also performed simultaneous PSF-fitting photometry on all images from the two datasets with the routine kitchen\_sync2 (Anderson 2020, in prep.). The final catalog includes 13,542 stars with at least one measurement in all the three filters, $F475W$, $F606W$, and $F814W$. This photometric catalog was calibrated to the VEGAMAG system by using the same zeropoints used to calibrate the DAOPHOT/ALLFRAME catalog.

This code was also used to run artificial star tests with the same input synthetic CMD used for the tests with DAOPHOT/ALLFRAME discussed above. We used a total of $\simeq$650,000 stars uniformly distributed on the field. The artificial stars are added to the image and measured by KS2 one at a time, in order of not to affect the original level of crowding.

\section{The Eridanus II CMD}\label{sec:cmd}

\begin{figure*}[!ht]
\centering\includegraphics[width = 0.8\textwidth]{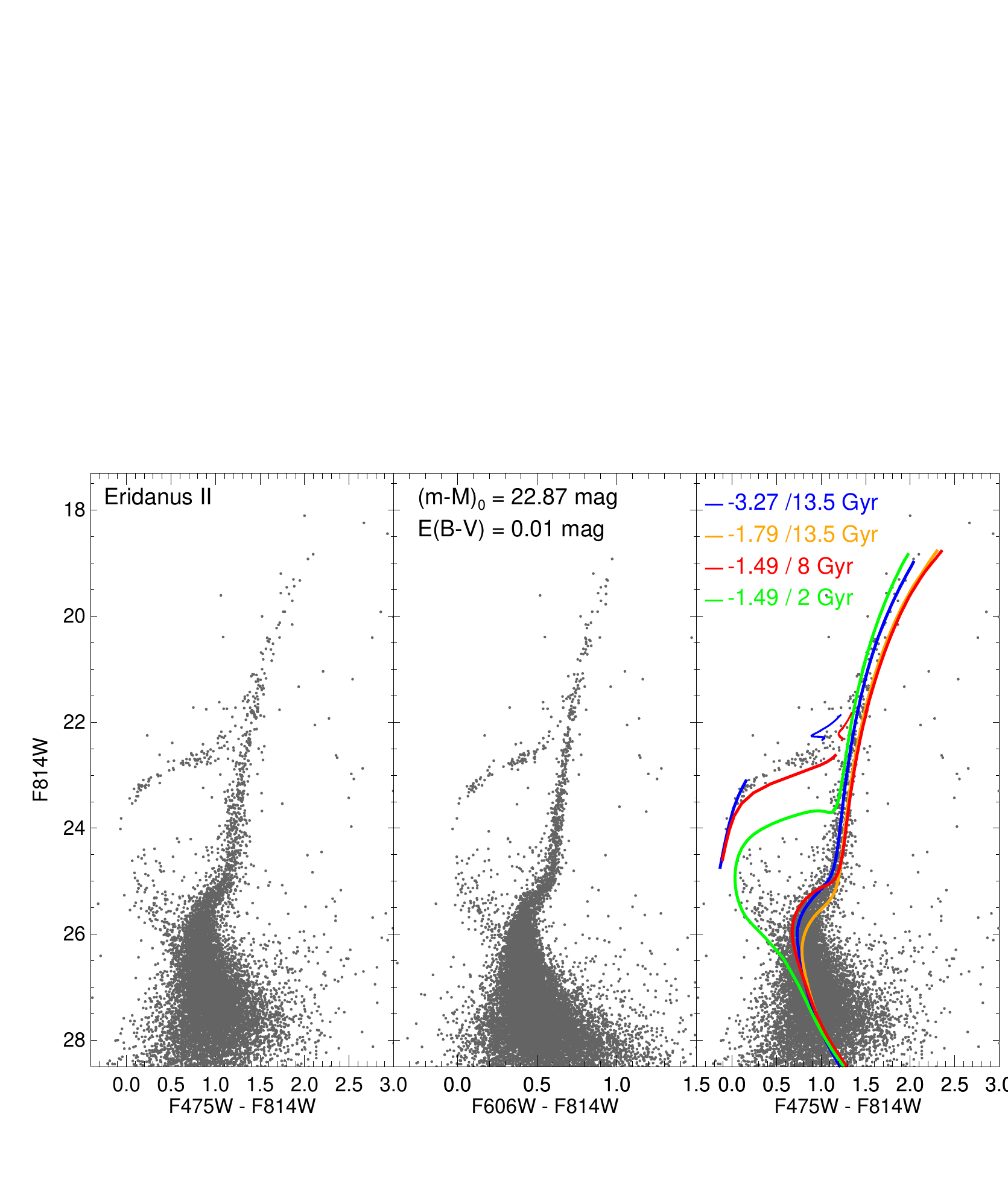}
\caption{Eri~II CMDs for the (F475W, F814W) and (F606W, F814W) band combinations. A number of isochrones and ZAHB from the BaSTI library of the ages and [Fe/H] labelled, and two He-burning tracks of 1.2 M$_\odot$ and [Fe/H] in the labelled color code, have been superimposed to the [(F475W-F814W), F814W] CMD (see text for details).} 
\label{iso_bi}
\end{figure*}

Figure~\ref{iso_bi} displays the [($F475W-F814W$), $F814W$] and [($F606W-F814W$), $F814W$] CMDs of the central region of Eri~II covered by the ACS data (left and central panels, respectively). The diagram spans about 10 magnitudes, from the tip of the RGB ($F814W\sim$ 18.5 mag) down to the limiting magnitude $F814W\sim$ 28.5 mag, which is about 3 magnitude fainter than the oMSTO. The plot discloses the  features typical of an old population. The bright part presents a remarkably steep RGB, suggestive of mostly metal-poor populations, while the horizontal-branch (HB) is well populated, on both the blue and the red part, in the colour range 0 $< F475W - F814W <$ 1.1 mag. A blue plume of stars brighter than the bulk of the main sequence stars has the typical appearance of a population of blue straggler stars (BSS), commonly observed  in globular clusters and dSph galaxies. In fact, the logarithmic frequency of BSS in Eri II, calculated identically as in \citet{Monelli2012bss} is 0.153, which taking into account the absolute magnitude M$_V = -7.1$ of Eri II, places this galaxy right in the relationship defined by LG dwarfs containing a sequence of BSS in the CMD \citep[see Figure 7 in][]{Monelli2012bss}.  

In the right panel of Figure~\ref{iso_bi}, some isochrones selected from the BaSTI database \citep{Pietrinferni2004} are superimposed to the CMD. We adopted the distance modulus derived from the properties of the RR Lyrae stars  (Mart\'\i nez-V\'azquez et al. in prep), (m-M)$_0=22.87\pm0.10$ and extinction A$_{F475W}$ = 0.033 mag, and A$_{F814W}$ = 0.015 mag from \citet{Schlafly2011}. The figure discloses that a simple comparison with isochrones is not enough to assess the age spread of the dominant  population of Eri II. In fact, the plot shows that the morphology of the main sequence and the RGB can be explained by different combinations of age and metallicity. A very old (13.5 Gyr) and extremely metal-poor  (Z=10$^{-5}$ or [Fe/H]=-3.27, blue line) isochrone matches the blue envelope of the CMD, from the oMSTO to the sub-giant branch and up to the tip of the RGB. An isochrone of the same age and 30 times more metal-rich (Z=0.0003, [Fe/H]=-1.79, orange line) nicely brackets the red envelope of the same features. However, a much  younger (8 Gyr) and even more metal-rich (Z=0.0006, [Fe/H]=-1.49, red line) represents well the turnoff and subgiant-branch shape, and the red part of the RGB.  The plume of stars bluer and brighter than the oMSTO is well matched by a 2 Gyr old isochrone (green). However, as mentioned above, the location of these blue and bright main sequence stars is also consistent with a population of BSS.  Finally, two zero-age HB (ZAHB) loci, and two He-burning tracks of 1.2 M$_\odot$, for the lowest and highest metallicity considered in this section, are also displayed. The comparison between the ZAHBs and the observed HB stellar distribution shows that the blue HB stars can be accommodated as belonging to the most metal-poor stellar population in the galaxy while the red HB are consistent with a more metal-rich stellar component, with [Fe/H] up to $-1.49$, consistent with the maximum spectroscopically observed metallicity \citep[][see also Section \ref{sec:sfh_comment}]{Li2017}. However, since the ZAHB location of a star strongly depends on its total mass -- the larger the mass the redder the ZAHB location -- some of the reddest and brightest stars in the HB could be progeny of the BSS population mentioned before (see the position of the He-burning tracks).

In summary, the main sequence locus observed in the Eri~II CMD, can easily accommodate a range of ages of up to 5-6 Gyr, and a range of metallicities of over 1.5 dex, while a similar range of metallicities can also explain the HB morphology. The quantitative analysis of the SFH presented in Section~\ref{sec:sfh_derivation} will allow us to disclose the actual stellar composition of the galaxy: whether star formation extended for a few Gyr, and whether the blue plume is consistent with a BSS population or, on the contrary, a young population is necessary to explain it.

\section{Star Formation History derivation}\label{sec:sfh_derivation}

The SFH of Eri~II has been derived through the comparison of the observed CMDs with synthetic CMDs to extract the combination of simple stellar populations (SSP, that is, synthetic populations with small ranges of ages and metallicities) which provide the best fit. The SFH provides a detailed and robust determination of the star formation rate (SFR) and metallicity Z as a function of time. In order to establish the robustness of the SFH determination, two different stellar evolution model sets (BaSTI and PARSEC) and two independent SFH derivation procedures (TheStorm and SFERA) have been used. Additionally, the SFH determination with TheStorm and BaSTI models has been performed using the two independent photometry catalogues obtained with DAOPHOT/ALLFRAME and ePSF as described in Section \ref{sec:data}. Finally, to further constrain the actual duration of the old star formation burst in Eri~II, a number of tests with mock stellar populations have been performed.  

\subsection{SFH with TheStorm}\label{storm_derivation}

The majority of the solutions for the SFH that will be discussed in this paper are derived using the code TheStorm presented in \citet{Bernard2015letter, Bernard2018}, which closely follows the procedures adopted and discussed in detail in previous papers \citep{iacpop, Monelli2010sfhcetus,Hidalgo2011sfhlgs3}.
This code uses a 'mother' synthetic CMD from which the SSP are extracted. We calculated three mother synthetic CMDs adopting the set of BaSTI models accounting for core convective overshooting during the core H-burning stage and mass loss during the RGB evolution\footnote{It is worth noting that the choice between canonical and overshooting models is quite irrelevant for a galaxy such as Eri~II whose bulk stellar population is old.}. We have alternatively adopted both the solar-scaled release of the BaSTI library \citep{Pietrinferni2004} and the $\alpha$-enhanced one (\citealt{Pietrinferni2006}, see Table~\ref{models}). 

We note that the version of the BaSTI stellar evolution library adopted in the present work does not take into account the occurrence of atomic diffusion in the evolutionary computations - although it has been considered in the calibration of the corresponding Solar Standard Model \cite[see][for a detailed discussion on this issue]{Pietrinferni2004}. The impact of atomic diffusion is to decrease the age of the oldest stellar populations by about 1 Gyr (the exact value depending on the absolute age and the stellar metallicity)  with respect to an age dating procedure based on non-diffusive stellar models \citep[see e.g.][for a detailed discussion on this topic]{Cassisi_Salaris2013}. In order to address this issue in detail for the current problem, we have compared suitable stellar models computed on purpose that alternatively account or not for atomic diffusion. We have verified that, for a low metallicity such that of Eri II, a 13.5 Gyr diffusive isochrone mimicks a non-diffusive one about 0.6-0.7 Gyr older. Therefore, in order to account for this characteristic of the BaSTI models, in computing the synthetic CMDs we included star of ages up to 14 Gyr old (see below).

For obtaining the synthetic CMDs, we used a customized version of the Stellar Population Synthesis Program available at the BaSTI web
page\footnote{http://basti.oa-teramo.inaf.it/}, which allows the computation of synthetic CMDs with flat age and metallicity distributions within given age and metallicity ranges. This code has been discussed in \cite{Cordier2007} and here we will just highlight that a binary star population is generated according to  a binary fraction $\beta$ and minimum mass ratio $q$ specified by the user, with the mass of the secondary selected randomly following the recipe by \cite{Woo2003}. The current version does not account for any stochastic variation of the mass loss efficiency during the RGB evolutionary stage. The effect of the occurrence of mass loss is accounted for by assuming a constant value for the free parameter $\eta=0.4$ in the Reimers formula in the stellar model computations \citep[see,][for a detailed discussion on this issue]{Pietrinferni2004}. Finally the bolometric correction library for the ACS photometric system adopted in the present work is that presented by \cite{Bedin2005}.

Common to all mother synthetic CMDs (with $5\times10^7$ stars) are a flat SFR between ages 14 and 2 Gyr (model stars younger than this age have not been included because there is no evidence that they are present in Eri~II), and the \citet{Kroupa1993} initial mass function (IMF). We have computed three synthetic CMDs with the stellar evolution models set (solar-scaled or $\alpha$-enhanced) and
characteristics of the binary star population as detailed in Table \ref{models}. A range in metallicity consistent with that determined in \citet{Li2017} was adopted. This required the use of the lowest metallicity stellar models available in the BaSTI database (Z=0.0001), with an upper Z=0.0004 for the scaled-solar models and Z=0.001 for the $\alpha$-enhanced ones. The metallicity distribution in the mother synthetic CMD is also flat in the whole metallicity range above, that is, no other constrains on the metallicity distribution of the galaxy are imposed. 

\begin{deluxetable}{ccccccc}
\tabletypesize{\footnotesize}
\tablecaption{Parameters adopted for the synthetic CMDs \label{models}}
\tablecolumns{6}
\tablewidth{0pt}
\tablehead{
\colhead{Library} & \colhead{Name} & \colhead{Set$^\tablenotemark{a}$} & \colhead{Diffusion} & IMF & \colhead{$\beta$} & \colhead{q}
}
\startdata
BaSTI & aeb05 & ae & NO & Kroupa (1993) & 0.5 & 0.4 \\
BaSTI & ssb05 & ss & NO & Kroupa (1993) & 0.5 & 0.4 \\
BaSTI & ssb07 & ss & NO & Kroupa (1993) & 0.7 & 0.1 \\
PARSEC &  parsec  & ss & YES & Kroupa (2001) & 0.3 & 0.0 \\
\enddata
\tablenotetext{a}{Model set: ae: $\alpha$-enhanced chemical composition; ss: solar-scaled chemical composition}
\end{deluxetable}

Observational errors are simulated in the mother synthetic CMD to create a model CMD that  is directly comparable to the observations. A C++ parallel code called DisPar which follows the strategy detailed in \citet{Gallart1996a} and discussed in \citet{Ruiz-Lara2020MNRAStmp} has been used. It takes into account all the observational effects as derived through the artificial stars tests, using an empirical approach. This is a fundamental step because the distribution of stars near the oMSTO is modified substantially due to the observational errors, and their realistic simulation is fundamental to derive a reliable and detailed SFH at the oldest age. Even though DisPAR can take into account possible spatial variations of the errors derived with the artificial stars tests when simulating them in the mother CMD, we have not used this feature in this case, as the crowding is low across the image, and results in very uniform observational errors.

The model CMDs have been divided in SSPs using a number of age and metallicity bins. Age bins are of 0.5 Gyr width in the whole age range considered (14.0 to 2.0 Gyr ago). The choice of an age bin of width of only 0.5 Gyr even at the oldest ages was made after some testing that showed that wider bins did slightly artificially widen the old star formation burst derived for the galaxy. To sample the distribution of stars in the CMD we have defined a number of bundles \citep[a collection of non-overlapping regions in color-magnitude space, see][]{Monelli2010sfhcetus}, which are shown in Fig. \ref{residuals} for the [(F475W-F814W), F814W] CMD (a similar distribution of bundles was adopted for the[(F606W-F814W), F814W] CMD). Each bundle was divided in boxes of different size, as indicated in the Figure inset. In this case, and after some testing \citep[see also][]{Ruiz-Lara2020MNRAStmp}, we decided to include the whole CMD above the 80\% completeness limit within the set of bundles, not excluding regions sampling stars in advanced evolutionary phases which may be less well described by current stellar evolution models. The influence in the fit of different regions of the CMD has been instead modulated by the different sizes of the boxes in each bundle, with smaller, thus more numerous boxes in the more populated, best theoretically understood \citep {Gallart2005ARAA} main sequence near the oMSTO.

The best-fitting SFH is determined by finding the amplitudes of the linear combination of SSP CMDs that best match the observed CMD. No a-priori age-metallicity relation is adopted. The number of observed and synthetic stars from each SSP, counted in each CMD box, serves as the input to the minimization routine, which measures the goodness of fit using a Poisson equivalent of $\chi^2$ \citep{Dolphin2002}. The minimization process is repeated several times, each one shifting the observed CMD in  steps in color and magnitude with respect to the synthetic CMDs, in order to account for uncertainties in photometric zeropoints, distance, and mean reddening. Once the shift that leads to the best solution is located, another set of solutions are calculated by shifting the bins used to sample the CMD, and also the age-metallicity bins used to define the SSP in the case of the model CMD. The SFH represented in Fig.~\ref{sfhplot_storm} is the result of averaging 25 such solutions. The total uncertainties are assumed to be a combination (in quadrature) of the uncertainties due to the effect of binning in the colour-magnitude and age-metallicity planes, and those due to the effect of statistical sampling in the observed CMD \citep[see][for details]{iacpop}.

\subsection{SFH with SFERA}\label{cignoni_derivation}

The SFH of Eri~II was also recovered using the hybrid-genetic algorithm SFERA (Star Formation Evolution Recovery Algorithm). 
The details of this approach are described in \citet{Cignoni2015} and here we provide only a short description. In this procedure, a library of ``basic" synthetic CMDs (similar to the SSP defined above) is generated using the latest PARSEC-COLIBRI isochrones (PAdova and TRieste Stellar Evolution Code version 1.2S plus COLIBRI code for AGB thermal pulses; \citealt{Bressan2012,Tang2014,Marigo2017}). Each basic synthetic CMD is a Monte Carlo realization with constant SFR (in a given range of ages), \cite{Kroupa2001} IMF (between 0.1 and 300 M$_{\odot}$), fixed metallicity ($\pm\, 0.05$ dex). In this work we adopted a logarithmic time binning of 0.01 in the interval $\log({\mathrm{age}}) = 9.90-10.13$, and a time binning 0.1 in the interval $\log({\mathrm{age}}) = 9.50-9.90$. Concerning metallicity, we allowed the code to use metallicities\footnote{We adopt the approximation [M/H]$ \simeq \log(Z/Z_{\odot})$, with $Z_{\odot}=0.0152$.} between $[M/H] = -2.2$ and $-1.0$ with a resolution of 0.1 dex. The last step concerns the binary population: 30\% of synthetic stars are considered to be part of unresolved binary systems and their flux is combined with a companion, whose mass is a random fraction between 0
and 1 of the primary.

Once the library is created, the models are convolved with photometric errors and incompleteness as derived from the artificial star tests performed on the real images and described in section \ref{daophot}. The linear combination of basic CMDs which minimizes the residuals from the observational CMD (in terms of Poissonian likelihood) is searched with a hybrid-genetic algorithm, combining the exploration ability of a genetic algorithm and the exploitation ability of a local search. 

\section{The SFH of Eridanus II}\label{sec:sfh_comment}

The SFH of Eridanus II is displayed in Figures~\ref{sfhplot_storm}
and~\ref{sfhplot_cignoni}. In Figure~\ref{sfhplot_storm}, the results by TheStorm DAOPHOT CMD fit are shown. Lines of different colors indicate the different solutions depending on the model CMD (see Table~\ref{models}) and the band combination used, as indicated in the labels (F475W or F606W for the bluest band).  The top panel shows SFR(t), the SFR as a function of time, which is characterized by a strong, narrow episode of star formation at the oldest possible age, and lasting $\simeq 0.5$ Gyr (half width at half maximum: this is the maximum possible duration of this early period of star formation; we try to constrain its intrinsic duration in Section~\ref{sec:mock}). After this main star forming period, a second period of possible extremely low level star formation activity, lasting until $\simeq$ 9 Gyr ago, is found in our solutions. This feature is somewhat more prominent in the solutions obtained from the (F606W, F814W) CMDs, in which the SFR(t) displays a second small maximum of low significance around 12 Gyr ago. The lower panel displays the cumulative mass fraction, which shows that the star formation after the main period ending $\simeq$ 13 Gyr ago could amount to up to $\simeq$ 20\% of the stellar mass formed by the galaxy. However, note that, in all cases, this possible low level SFR is compatible with null star formation after the main burst. The SFHs recovered also contain information on the age-metallicity relation and metallicity distribution that provide the best fit to the CMDs. In this case, since the range of age is very small for a meaningful exploration of the age-metallicity relation, we will consider the recovered metallicity distribution. Figure \ref{ChemHisto} displays a number of metallicity histograms derived from the solution CMDs constructed from the SFHs, compared with the distribution of metallicities measured by \citet{Li2017} for 12 Eri II stars. Note the good agreement, particularly for the solutions with the $\alpha$-enhanced models, in spite of the low number of spectroscopically measured stars.  

\begin{figure}[h!] 
\centering\includegraphics[width=0.5\textwidth]{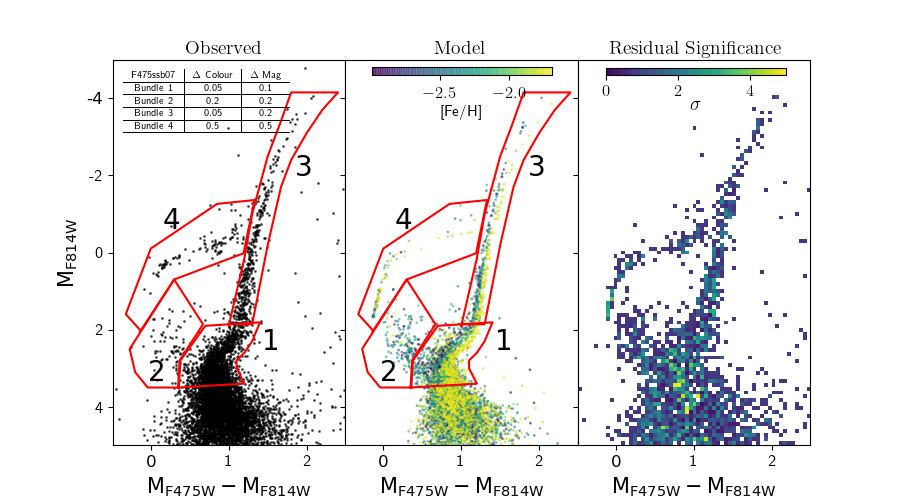} 
\caption{Left panel: Observed [(F475W-F814W), F814W] CMD of EriII.  The bundles used for the SFH derivation have been superimposed, and the size of the boxes in each bundle are given in the inset. Middle panel: Best fit CMD corresponding to the F475ssb07 solution. The synthetic stars in this CMD have been color-coded according to their [Fe/H], as indicated in the upper color bar.  Right panel: Significance of the residuals in units of Poisson $\sigma$.}
\label{residuals}
\end{figure}

\begin{figure}[h!]
\centering\includegraphics[width=0.45\textwidth]{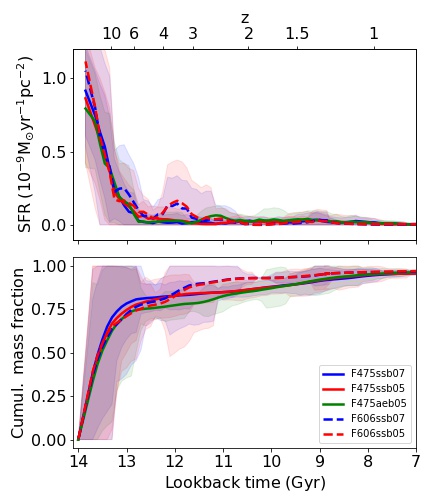} 
\caption{SFH of Eri~II derived with TheStorm. The upper panel shows the SFR while the lower panel shows the cumulative mass fraction as a function of the look-back time. Different lines indicate different solutions combining the DAOPHOT photometry of the two datasets (F475W+F814W or F606W+F814W), different BaSTI stellar evolution model sets (ss or ae), and binary stellar population properties (binary fraction $\beta$=0.7 or 0.5). See Table~\ref{models} for details.}
\label{sfhplot_storm}
\end{figure}

\begin{figure}[h!]
\centering\includegraphics[width=0.45\textwidth]{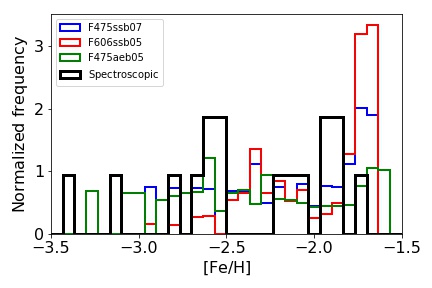} 
\caption{Normalized metallicity histograms derived for Eri~II from a number of DAOPHOT/TheStorm solutions in Figure~\ref{sfhplot_storm}, compared with the metallicities derived by \citet{Li2017}.}
\label{ChemHisto}
\end{figure}

\begin{figure}[h!] 
\centering\includegraphics[width=0.45\textwidth]{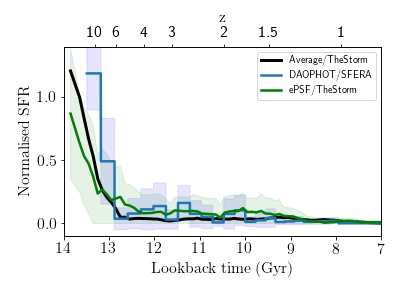} 
\caption{Comparison between the average of the SFR(t) solutions from the DAOPHOT (F475W,F814W) CMD obtained with TheStorm displayed in Figure~\ref{sfhplot_storm}, one solution obtained with TheStorm using the ePSF (F475W,F814W) CMD, and with a solution obtained with SFERA and a DAOPHOT (F475W,F814W) CMD. The SFR(t) plots have been normalized by the area below the curve. Note the excellent agreement between all the solutions.}
\label{sfhplot_cignoni}
\end{figure}

Figure~\ref{residuals} shows the observed CMD (left panel), an example solution CMD (middle panel, color coded with [Fe/H], see below), and the significance of the residuals in units of Poisson $\sigma$ (right panel). It can be seen that the resulting solution CMD is an overall good realization of the observed CMD: no significant structure is observed in the residual plot, and most CMD bins show residuals under 2$\sigma$. One area of the CMD where the discrepancy between observed and solution CMD is particularly apparent is in the HB, which is substantially more extended and more populated toward the blue side in the model. This can be understood by the fact that the RGB mass loss efficiency is parameterized in a simple way and not optimized to provide the best possible match with the data (see Section~\ref{sec:sfh_derivation}; if $\eta=0.2$ in the Reimers formula had been used to parameterize the mass loss along the RGB, the qualitative agreement between the observed and the model RGB would have been likely better, as a smaller mass loss results in a HB less extended to the blue). In spite of this, in the middle panel it can be seen that the HB well populated from the blue to the red can be explained by the wide range of metallicity (spanning over 2 dex) in the galaxy. In fact, from the solution CMDs, we calculated that $\sim$ 5 stars can be expected for the evolved progeny of the bright main sequence stars, and thus their contribution to the star counts in the red part of the HB is small. This panel also shows that the bright main sequence (that we tentatively interpreted as a BSS population in Section \ref{sec:cmd}) is best represented by stars in the whole metallicity range, with numerous low metallicity stars. This reinforces the conclusion advanced in that section that this sequence is composed by BSS stars and not by stars younger (and thus likely more metal rich) than the bulk of the Eri II stars.

Figure~\ref{sfhplot_cignoni} compares the average of the three F475W TheStorm/DAOPHOT solutions (black) shown in Figure~\ref{sfhplot_storm}, with i) one solution obtained with TheStorm using the ePSF (F475W,F814W) CMD (green), and ii) a solution obtained with SFERA and a DAOPHOT (F475W,F814W) CMD (blue). The SFR(t) have been normalized to the total area below the curve. Remarkably, the independent solutions obtained with different SFH derivation codes adopting different stellar evolution libraries and stellar populations parameters, such as the IMF or the binary stars characteristics, present the same key features for the Eri~II SFH. Also the solutions with TheStorm for different photometry sets are consistent. In the rest of the paper, and for simplicity, we will adopt and discuss in more detail the DAOPHOT/TheStorm solution. We will also perform most of the analysis in relation to the robustness of the features in the Eri~II SFH using TheStorm. 

\subsection{Age accuracy and precision at old ages} \label{sec:mock}

In order to assess the robustness of our conclusions regarding the key features in the  Eri~II SFH, and to further constrain the actual intrinsic duration of the main star formation episode, and the precision in dating this burst, in the following sections, we will perform a number of tests with simulated CMDs of stellar populations of known age characteristics.

\subsubsection{Synthetic mock galaxies}

\begin{figure}[h!]
\centering\includegraphics[width=0.5\textwidth]{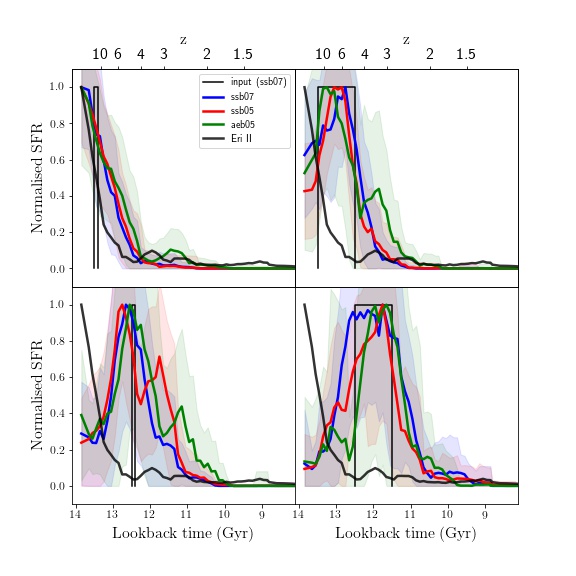}
\caption{Results of the test with synthetic mock galaxies described in the text. The black thin lines indicate the SFR(t) of the mock galaxy, while the thick black line indicates the SFR(t) inferred for Eri~II (average of the F475aeb05 and F606ssb05 solutions). The colored lines indicate the SFR(t) derived for the corresponding mock galaxy, using the  (F475W, F814W) band combination and different mother synthetic CMDs (see Table~\ref{models}).}
\label{mocks}
\end{figure}

We have created four 'synthetic mock galaxies' from synthetic CMDs in which observational errors have been simulated using DisPar and the Eri~II artificial star tests. The synthetic CMDs have been calculated with the same synthesis program used to compute the mother synthetic CMDs, adopting the solar-scaled BaSTI stellar evolution library, a gaussian metallicity distribution with mean [Fe/H]=-2.38 and $\sigma_\mathrm{[Fe/H]}$=0.47
\citep[as observed for Eri~II by][]{Li2017} and a binary fraction $\beta=0.7$ with mass ratios q$ > $0.1. The difference between these four mock galaxies is the age distribution of their stellar populations: two of them were calculated with an age range of 100 Myr (between 13.5 and 13.4 Gyr old, and 12.5 and 12.4 Gyr old) and the other two with an age range of 1 Gyr (13.5-12.5 Gyr old, and 12.5-11.5 Gyr old). The shape of these input SFR(t) are represented as thin black lines in Figure~\ref{mocks}. These tests were performed using both the (F475W,F814W) and (F606W,F814W) band combinations. After simulating the observational errors in these synthetic populations using the corresponding artificial star tests, a number of stars comparable with the observed CMDs of Eri~II was randomly selected to fully mimic the observed CMD properties. The CMDs of the four mock galaxies were treated in the same way as the observed Eri~II CMD to retrieve their SFH with TheStorm. The same three model CMDs used to retrieve the SFH of Eri~II (see Table~\ref{models}) were also used for the mock galaxies, in order to mimic the fact that the assumptions that go into the model CMD creation (e.g. stellar evolution, binaries) are different to the {\it parameters} of the actual galaxy. The recovered SFR(t) for the mock galaxies are displayed in Figure~\ref{mocks} (colored lines), together with the Eri~II SFR(t) (thick black solid line). The results for the mocks calculated with the two band combinations are basically identical, and in the Figure we show those for the (F475W, F814W) band combination, while for the Eri II solution we show the average of the F475aeb05 and F606ssb05 solutions.

 It can be seen that the recovered SFR(t) do not depend strongly on the adopted model CMD (with slightly worse recovery for 'aeb05', that is, when the stellar evolution model set used to compute the mother CMD not matches the one used to calculate the mock galaxy: $\alpha$-enhanced for the model CMD and solar scaled for the mock galaxy), and that the age of the population is well recovered in all cases. The true width of the star formation episode is not well recovered in the 100 Myr case. This indicates that we cannot resolve a narrow star formation event occurred $\simeq$ 13.5-12.5 Gyr ago with these data. However, note that only the mock galaxy with the {\it narrow and old age range (13.5-13.4 Gyr)} results in a recovered SFH similar to the one derived for Eri~II, which is actually in the narrow side of the three SFR(t) derived for the mock. This indicates that the Eri~II main star forming epoch occurred at the earliest possible age, and might have been very narrow, possibly not lasting longer than 100 Myr. On the other hand, for the oldest, best fitting mock SFHs, the period of very low star formation activity extended to $\simeq$ 9 Gyr ago that is observed in the Eri~II SFR(t) is not observed, except in the case in which there is a mismatch between the parameters used to compute the model CMD and the mock galaxy (green lines). In this case, a local maximum of low intensity, similar to the one observed in the original F606W solutions, is recovered at an look back time of $\simeq$ 11.5 Gyr ago. This is an indication that these low level features in the SFR(t) are likely spurious. 

\subsubsection{Retrieving the SFH of a M92 mock}

\begin{figure}[h!]
\centering\includegraphics[width=0.5\textwidth]{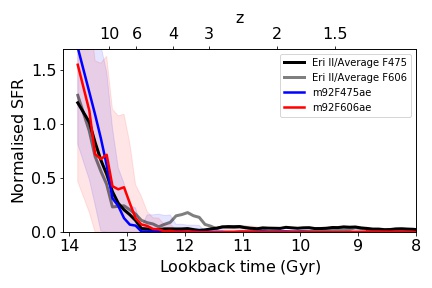}
\caption{Comparison between the SFR(t) obtained for the M92 mock described in the text, and for Eri~II. The blue and red line refer to the mocks constructed with the [F475W,F814W] and [F606W,F814W] M92 data, respectively. The gray and black lines represent the average of the F475W and F606W SFR(t) solutions for Eri~II shown in Figure \ref{sfhplot_storm}.} 
\label{m92}
\end{figure}

It could be argued that the tests with mock galaxies presented in the previous section are idealized cases in which the same stellar evolution model set has been used to calculate the input mock stellar population and the model synthetic CMD used to retrieve its SFH. We have performed a further test with a mock stellar population created using the CMD of the old,  metal poor ([Fe/H]~=~--2.3) Milky Way globular cluster M92  \citep[NGC6341]{DiCecco2010, VandenBerg2016}, as observed with the ACS under HST programs ID. 10505 (1 orbit in each F475W, F814W) and ID. 10775 (1 orbit in each F606W, F814W). Since these data are much deeper in the absolute magnitude planes than the Eri~II CMD, we have assumed that the photometric errors are negligible, and degraded the M92 CMDs to the photometric quality of the Eri~II CMDs, by simulating in them the Eri~II observational errors using the Eri~II artificial stars tests. We also selected a similar number of stars as in the Eri~II observed CMDs. After that, the SFH was obtained identically as for Eri II. Figure~\ref{m92} shows the SFR(t) obtained for the M92 mock, using either the [F475W, F814W] (blue line) or the [F606W, F814W] (red line) CMDs and adopting the model synthetic CMD computed with the $\alpha$-enhanced model set and $b=0.5$, compared to the Eri~II SFH. The width of the main star forming period of Eri~II is basically identical to that derived for M92, thus reinforcing the conclusion in the previous section that the early Eri II star formation burst must have had a very narrow timespan, possibly even narrower than 100 Myr.  In this case, the period of star formation extended to $\simeq$ 9 Gyr ago is clearly not found. 

\section{Discussion} \label{discussion}

 The SFHs of Eri II presented in Section \ref{sec:sfh_comment}, obtained using different photometry tables and CMD-fitting methods, and constrained further with tests using mock stellar populations all agree in that Eri~II has formed the bulk of its stars in a very early and extremely short (possibly shorter than 100 Myr) star formation burst.  As discussed in the Introduction, three main mechanisms are usually debated as potentially able to remove gas and halt star formation in small galaxy halos, namely i) ram-pressure stripping and tidal interactions with a nearby larger galaxy, ii) cosmic reionization, and iii) SNe feedback associated to the early star formation in the dwarf galaxy itself. Since we have a quantitative determination of the amount of mass turned into stars in Eri~II early star formation burst, and stringent constraints on the duration of this burst, we will investigate whether Eri II could have removed its own gaseous component at early times just by stellar feedback caused by supernovae (SNe) explosions. For this, we follow the methodology by \citet{Bermejo2018}. We first integrate the SFH of Eri~II and correct the result by the missing stellar component in the $HST$/ACS observations, which cover $\sim$ 55\% of the galaxy, taking into account the galaxies stellar profile derived by \citep{Crnojevic2016}. We obtain that the total mass transformed into stars is $\sim 1.9 \times 10^5 M_\odot$. Assuming the IMF by \citet{Kroupa2001} and a stellar mass threshold of 6.5 $M_\odot$ for explosions of SN~II, we get that Eri~II had $\sim$ 2700 SNe events, which considering a typical energy $\sim 10^{51}$ erg per SN would have injected an energy $\sim 2.7 \times $10$^{54}$ erg to the medium.  We note that when adopting a higher mass threshold, like 8 $M_\odot$ for the SN~II explosions, the energy budget would be $\sim 75\%$ the value obtained for 6.5 $M_\odot$.

To evaluate whether this energy could or could not push away the gas, we assume that all the SNe events in Eri~II happened before $z \sim 6$ and calculate the competing gravitational potential of the dark matter halo at this redshift. According to \citet{Woo2008}, we consider that this kind of dwarf galaxies with a very {\it fast} SFH \citep{Gallart2015} present a stellar mass-to-light ratio close to $\sim 2$, and we derive the present-day stellar mass from the luminosity by \citet{Crnojevic2016}. We link the resulting stellar mass ($\sim 1.1 \times 10^5 M_{\odot}$) to the $z=0$ virial halo mass using the abundance-matching (AM) relations by \citealt{Brook2014} (B14)  and \citealt{Moster2013} (M13). We evolve these virial halo masses back to $z=6$  following \citet{Fakhouri2010} and we obtain halo masses of $3.4 \times 10^8 M_{\odot}$ and $2.1 \times 10^8 M_{\odot}$ for the B14 and M13 AM relations, respectively. Then, we derive the gravitational potential of the dark matter halo and relate it to the gaseous component accounting for the baryon fraction ($f_b \sim 1/6$). We finally obtain that the competing gravitational potential is $\Delta W_{\rm gas}^{z = 6}/2 = 4.8 \times 10^{52}$ erg and $\Delta W_{\rm gas}^{z = 6} /2 = 2.2 \times 10^{52}$ erg for the B14 and M13 AM relations, respectively. This means that Eri II could have removed its gaseous component with a coupling efficiency of the SNe energy $\epsilon \lesssim 2 \%$, which is well compatible with the maximum values of this parameter in the literature; for example in \citet{Bermejo2018} the coupling efficiency is constrained to be smaller than $\sim 10\%$.

The above estimates are of course an order of magnitude calculation since i) we are assuming that all SNe explode at once, thus producing a much powerful explosion than if they were spread out in time, and ii) we are relying on a dark matter halo mass derived from abundance matching relations.
However, the very low star formation rate of Eri~II, $\langle\Psi \rangle \approx 10^{-3} M_{\odot}/$yr, is perfectly consistent with what is predicted for inefficiently star forming mini-halos with relatively small virial halo mass, $\approx 10^7 M_{\odot}$, which can rapidly lose all their gas thanks to SN feedback and then are permanently switched off because reionization is preventing further gas accretion \citep[e.g.][Rossi et al. submitted]{SalvadoriFerrara2009}. In conclusion, the short history of star formation of Eri~II, lasting $<100$~Myr combined with its low star formation rate, strongly support the idea that we are observing the living relic of one of the first star forming ${\rm H_2}$-cooling minihalos, which hosted the first stars.

In absence of quantitative estimates of the actual early star formation rate (in units of solar masses per year), and thus of the SNe energy ejected to the interstellar medium, and noting the very similar quenching times in a number of UFD, previous works \citep[e.g.][]{Brown2012, Brown2014} favoured the conclusion that an outside global influence such as cosmic reionization was the most plausible cause of the synchronized quenching.  The remaining candidates, tidal interactions or ram-pressure stripping, would result in quenching times that depend on the accretion time of each dwarf galaxy into the Milky Way. In a paper that was accepted and posted in ArXiv after this one was submitted, \citet{Simon2020arXiv} analyzed the same Eri II ($F606W-F814W$, $F814W$) CMD with the methodology used in \citet{Brown2014}. They reach a conclusion on the age and duration of the star formation in Eri II that is totally compatible with our scenario. The main difference between our conclusion and theirs is in the interpretation of the results. While they focus their discussion on possible reionization and tidal effects, and conclude that the former is the most plausible explanation for the early quenching of Eri II, our quantitative estimate of the star formation rate allows us to argue that SN feedback is a main driver of the star formation quenching in Eri II.
Note that if the very early star formation events in UFDs as reported in \citet{Brown2014} were intense enough to result in sufficiently large feedback energy to remove the gas as in the case of Eri~II, they would also result in apparently synchronized quenching. 

\begin{deluxetable}{lccc}
\tablecaption{Properties of Eridanus II compared to Leo T\label{eri_leo}}
\tablecolumns{4}
\tablehead{
\colhead{Parameter} & \colhead{Eri II} & \colhead{Leo T} & \colhead{ref(LeoT)$^\tablenotemark{a}$}
}
\startdata
M$_V$ (mag) & -7.1$\pm$ 0.3 & -7.1  & (1) \\
r$_h$ (pc) & 252 & 166 & (1)\\
D$_{MW}$ (Kpc) & 375 & 409 & (2) \\
$v_{GSR}$ (km s$^{-1}$ & -66.6  & -58.4 & (3) \\
$\sigma_{v_{hel}}$ (km s$^{-1}$) & 6.9  & 7.5 & (3)\\
M$_{HI}$ (M$_{\odot}$) &  $< 2.8\times10^3$ & 2.8$\times 10^5$ & (4)\\
M$_{1\over2}$ (M$_{\odot}$) &  1.2$^{+0.4}_{-0.3}\times10^7$ & 8.2 $\pm$ 3.6 $\times10^7$ & (3) \\
M/L (M$_{\odot}$/L$_{\odot}$) & 420$^{+210}_{-140}$ & 138 $\pm$ 71 & (3)\\
$[Fe/H]$ & -2.38 $\pm$ 0.13 & -2.29 & (3)\\
$\sigma_{[Fe/H]}$ & 0.47  & 0.35 & (3)\\
\enddata
\tablenotetext{a}{The corresponding references for Eri II are given in Table \ref{basic_data}}
\tablecomments{(1) \citet{Irwin2007LeoT}; (2) \citet{Clementini2012}; (3) \citet{SimonGeha2007} (4) \citet{RyanWeber2008}}
\end{deluxetable}

That reionization may not have been the main or only cause of the star formation quenching in Eri II is supported by the fact that another UFD galaxy of similar properties such as Leo T (in terms of mass, metallicity, distance and velocity with respect to the Milky Way; see Table~\ref{eri_leo}) has been able to retain gas and sustain star formation until (almost) the present time \citep[][Surot et al., in prep; other gas-rich UFDs are discussed by \citealt{McQuinn2015LeoP_ACS} and \citealt{Janesh2019_HI_UFD}]{Clementini2012, Weisz2012LeoT}. The available SFHs of Leo T all agree on a low early star formation rate, $\sim$10$^{-5} M_\odot yr^{-1}$, around two orders or magnitude lower than the one we have inferred for Eri II, that would have not been able to quench it by SNe feedback. However, apparently reionization was not able to quench it either: the highest temporal resolution SFR(t) for this galaxy published in both \citet{Clementini2012} and Surot et al. (in prep) show a continuous SFR(t) which increases substantially after 10 Gyr ago, with no apparent gap after reionization. 

What may be the physical reasons for the radically different SFHs for Eri II and Leo T, two otherwise quite similar dwarf galaxies (Table \ref{eri_leo})? \citet{Rey2020} use cosmological high resolution hydrodynamical simulations of field dwarf galaxies encompassing the Eri II and Leo T mass range to explore the mechanisms that allow some of these small galaxies, all quenched after reionization in their simulations, to reignite star formation at a late time. They find that these galaxies have been able to replenish their interstellar medium by slowly accreting gas until high enough densities are reached to self-shield the gas from the UV background. These models could explain the extended Leo T SFH, even if reionization would have affected it early, at a level not clearly detected by the measured SFHs. It is also possible that the coupling between reionization and SNe feedback results in enough energy for a complete removal of gas in Eri II, but not in Leo T. In fact, \citet{Jeon2017} find that while reionization plays an important role in inhibiting star formation, SNe feedback is crucial for the ultimate quenching of UFDs. In addition, UFDs have such shallow potential wells that the SF, and thus the feedback, might be very sensitive to the local environment and to the details of the ISM/CGM physics. For example, SNe feedback and outflow from a nearby dwarf galaxy can also quench star formation directly in these low mass systems\citep{Mina2020arXiv}. But one may still ask the further question of what is the actual origin of the different early SFR of these two galaxies. \citet{Gallart2015} use precise, time-resolved SFHs of Local Group classical dwarf galaxies derived from CMDs reaching the oMSTO and find that they can be assigned to two basic types: {\it fast dwarfs} that started their evolution with a dominant and short star formation episode, and {\it slow dwarfs} that formed a small fraction of their stars early and have continued star formation activity until, or almost, the present time. Given their SFHs, Eri II and Leo T would allow extending this definition to the UFD regime. \citet{Gallart2015} hypothesized that the distinction between fast and slow dwarfs reflected the characteristic density of the environment {\it where they formed}, with  fast dwarfs assembling quickly in high density environments where interactions triggering star formation were common, leading to a high SFR before reionization, and resulting in strong gas loss due to the combined effects of reionization and SNE feedback \citep[see also][]{BenitezLlambay2015}. On the contrary, slow dwarfs would have resulted from a slower mass assembly in lower density environments, implying lower initial SFR and ability to retain gas. The proper motions and orbit estimated for Eri II \citep{Fritz2018} do not totally preclude that Eri II was close to the Milky Way at early times \citep[see also][for a possibility of Eri II being a backsplash galaxy that has had close encounter with M31 before being accreted onto the Milky Way]{Buck2019}, while the 3D movement of Leo T in the Local Group is still highly uncertain. Therefore more precise information on the orbits of Eri II and Leo T  would help understand whether they can provide further support to the hypothesis by \citet{Gallart2015} on the origin of the dichotomy in the slow/fast dwarf galaxy types, and its extension to even lower masses.

Finally, the finding that the bulk population of Eri~II formed before 13 Gyr ago makes this galaxy one of the oldest among the faint dwarfs in the Local Group, and hence the ideal testbed for dark matter scenarios alternative to CDM that imply a delayed start of star formation in the lowest mass halos. This applies, for example, to Warm Dark Mater models (WDM) and the currently popular Fuzzy Dark Matter (FDM) models. \citet{Chau2017} have shown how stringent constraints on the thermal velocity of the WDM particle candidate can be imposed  based on how early the largest fraction of the stars in the lowest mass subhalos have formed. 
Their analysis did not include an object as old as Eri~II and yet, using less stringent star formation history data on some of the UFD, they were able to exclude WDM models with particle masses lower than 3 keV, as they would suppress excessively the collapse of halos below virial masses below $10^9 M_{\odot}$.  Eri~II's current mass estimate (within the half mass radius) is well below this limit \citep{Li2017}. Therefore, a similar analysis based don Eri~II could help to constrain WDM models with energies in the range 3-5 keV, which is the territory where  other probes, such as the Lyman alpha forest and gravitational lensing by halo substructure, are reaching their limits \citep[see][for a review]{Bullock2017ARAA}. The delayed SFH in low mass halos has been recently shown to be a distinctive feature of Fuzzy Dark Matter models because halo collapse is delayed due to the quantum pressure of the underlying Bose-Einstein condensate \citep{Mocz2020}. Of course for these applications determining the virial mass of Eri~II before infall is crucial. 

\section{Summary and final remarks} \label{summary}

We present new ACS/HST CMDs reaching the oMSTO with excellent photometric precision for the ultra-faint dwarf galaxy Eri II and derive a detailed and precise SFH for this galaxy by fitting these CMDs with synthetic CMDs. The photometry, obtained from two independent datasets providing (F475W, F814W) and (F606W, F814W) measurements, has been obtained using two commonly used programs, DAOPHOT IV/ALLFRAME \citep{Stetson94} and ePSF/KS2 \citep{Anderson2008ePSF}, and the CMD-fitting has been performed using two different codes and stellar evolution model sets: TheStorm \citep{Bernard2018} with the BaSTI stellar evolution models \citep{Pietrinferni2004, Pietrinferni2006} and SFERA \citep{Cignoni2015} with the PARSEC isochrones \citep{Bressan2012}. The SFHs derived using different combinations of photometry tables and CMD-fitting methods all agree in that Eri~II has formed the bulk of its stars in a very early (ending before 13 Gyr ago) and extremely short star formation burst: possibly shorter than 100 Myr, as indicated with tests with mock stellar populations, and compared with the SFH derived for the globular cluster M92. Though up to 20\% of its stars could have been formed in an extended period lasting a few Gyr after the main burst, the signal corresponding to this possible extended star formation is very low, and the error bars make the star formation after the main burst compatible with zero. Similarly, no significant signal is detected in the SFH from the blue plume extended to bright magnitudes above the old main sequence. This, together with the low metallicity inferred for these stars among other indicators, point to the fact that, like in many other dwarf galaxies \citep{Monelli2012bss}, this feature corresponds to a population of BSS. This result conclusively shows that Eri~II is an extremely old galaxy, with no young or intermediate-age star formation, as the earliest observations of this galaxy suggested \citep{Bechtol2015, Koposov2015a}.

From the quantitative determination of the amount of mass turned into stars in this early star formation burst ($\sim 2 \times 10^5 M_\odot$), we infer the number of SNe events and the corresponding energy injected into the interstellar medium. For reasonable estimates of the Eri II early virial halo mass (evolved from the current virial mass calculated using the present day stellar mass and published abundance matching relations) and appropriate values of the coupling efficiency of the SNe energy, we conclude that Eri II could have been quenched by SNe feedback alone.  This short history of star formation, combined with the low star formation rate, support the idea that Eri II is the relic of one of the first star forming ${\rm H_2}$-cooling minihalos, which hosted the first stars.

Our results cast doubts on the need to invoke cosmic reionization as the preferred or only explanation for the early quenching of old UFD galaxies. This conclusion is indirectly supported by the fact that another UFD galaxy of similar properties to Eri II such as Leo T (in terms of mass, metallicity, distance and velocity with respect to the Milky Way) has sustained star formation until just a few hundred million years ago. This pair of galaxies, particularly when their orbits will be available, are key to understand the origin of the dichotomy between fast and slow dwarfs \citep{Gallart2015}, extended to the lowest mass regime.

\acknowledgments

We thank the anonymous referee for a careful reading of the manuscript and many useful suggestions that have helped to improve the paper. This study was supported by NASA through grant GO-14224 from
the Space Telescope Science Institute, which is operated by AURA, Inc., under  NASA contract NAS5-26555. CG, MM, TRL, GB and FS acknowledge financial support through grants (AEI/FEDER, UE) AYA2017-89076-P, AYA2016-77237-C3-1-P (RAVET project) and AYA2015-63810-P, as well as by Ministerio de Ciencia, Innovaci\'on y Universidades (MCIU), through Juan de la Cierva - Formaci\'on grant (FJCI-2016-30342) and the State Budget and by Consejer\'\i a de Econom\'\i a, Industria, Comercio y Conocimiento of the Canary Islands Autonomous Community, through Regional Budget. SS acknowledges support from the ERC Starting Grant NEFERTITI H2020/808240 and from the PRIN-MIUR2017, ``The quest for the first stars", prot. n. 2017T4ARJ5. TRL acknowledges support from a Spinoza grant (NWO) awarded to A. Helmi. AM acknowledges financial support from FONDECYT Regular 1181797 and funding from the Max Planck Society through a Partner Group grant. M.C. acknowledges the support of INFN ”iniziativa specifica TAsP.


\facilities{HST(ACS)}





\textit{Software:} \verb|Astrodrizzle|, \verb|DAOPHOT| \citep{Stetson87,Stetson94}, \verb|TheStorm| \citep{Bernard2018}, \verb|numpy| \citep{numpy2020}, \verb|scipy| \citep{scipy2020}, \verb|astropy| \citep{astropy2018,astropy2013}, \verb|matplotlib| \citep{Hunter2007}.





\begin{thebibliography}{}
\expandafter\ifx\csname natexlab\endcsname\relax\def\natexlab#1{#1}\fi

\bibitem[{{Amorisco}(2017)}]{Amorisco2017Cluster}
{Amorisco}, N.~C. 2017, \apj, 844, 64

\bibitem[{{Anderson} \& {King}(2006)}]{AndersonKing2006}
{Anderson}, J., \& {King}, I.~R. 2006, {PSFs, Photometry, and Astronomy for the
  ACS/WFC}, Instrument Science Report ACS 2006-01, ,

\bibitem[{{Anderson} {et~al.}(2008){Anderson}, {King}, {Richer}, {Fahlman},
  {Hansen}, {Hurley}, {Kalirai}, {Rich}, \& {Stetson}}]{Anderson2008ePSF}
{Anderson}, J., {King}, I.~R., {Richer}, H.~B., {et~al.} 2008, \aj, 135, 2114

\bibitem[{{Aparicio} \& {Hidalgo}(2009)}]{iacpop}
{Aparicio}, A., \& {Hidalgo}, S.~L. 2009, \aj, 138, 558

\bibitem[{{Astropy Collaboration} {et~al.}(2013){Astropy Collaboration},
  {Robitaille}, {Tollerud}, {Greenfield}, {Droettboom}, {Bray}, {Aldcroft},
  {Davis}, {Ginsburg}, {Price-Whelan}, {Kerzendorf}, {Conley}, {Crighton},
  {Barbary}, {Muna}, {Ferguson}, {Grollier}, {Parikh}, {Nair}, {Unther},
  {Deil}, {Woillez}, {Conseil}, {Kramer}, {Turner}, {Singer}, {Fox}, {Weaver},
  {Zabalza}, {Edwards}, {Azalee Bostroem}, {Burke}, {Casey}, {Crawford},
  {Dencheva}, {Ely}, {Jenness}, {Labrie}, {Lim}, {Pierfederici}, {Pontzen},
  {Ptak}, {Refsdal}, {Servillat}, \& {Streicher}}]{astropy2013}
{Astropy Collaboration}, {Robitaille}, T.~P., {Tollerud}, E.~J., {et~al.} 2013,
  \aap, 558, A33

\bibitem[{{Astropy Collaboration} {et~al.}(2018){Astropy Collaboration},
  {Price-Whelan}, {Sip{\H{o}}cz}, {G{\"u}nther}, {Lim}, {Crawford}, {Conseil},
  {Shupe}, {Craig}, {Dencheva}, {Ginsburg}, {Vand erPlas}, {Bradley},
  {P{\'e}rez-Su{\'a}rez}, {de Val-Borro}, {Aldcroft}, {Cruz}, {Robitaille},
  {Tollerud}, {Ardelean}, {Babej}, {Bach}, {Bachetti}, {Bakanov}, {Bamford},
  {Barentsen}, {Barmby}, {Baumbach}, {Berry}, {Biscani}, {Boquien}, {Bostroem},
  {Bouma}, {Brammer}, {Bray}, {Breytenbach}, {Buddelmeijer}, {Burke},
  {Calderone}, {Cano Rodr{\'\i}guez}, {Cara}, {Cardoso}, {Cheedella}, {Copin},
  {Corrales}, {Crichton}, {D'Avella}, {Deil}, {Depagne}, {Dietrich}, {Donath},
  {Droettboom}, {Earl}, {Erben}, {Fabbro}, {Ferreira}, {Finethy}, {Fox},
  {Garrison}, {Gibbons}, {Goldstein}, {Gommers}, {Greco}, {Greenfield},
  {Groener}, {Grollier}, {Hagen}, {Hirst}, {Homeier}, {Horton}, {Hosseinzadeh},
  {Hu}, {Hunkeler}, {Ivezi{\'c}}, {Jain}, {Jenness}, {Kanarek}, {Kendrew},
  {Kern}, {Kerzendorf}, {Khvalko}, {King}, {Kirkby}, {Kulkarni}, {Kumar},
  {Lee}, {Lenz}, {Littlefair}, {Ma}, {Macleod}, {Mastropietro}, {McCully},
  {Montagnac}, {Morris}, {Mueller}, {Mumford}, {Muna}, {Murphy}, {Nelson},
  {Nguyen}, {Ninan}, {N{\"o}the}, {Ogaz}, {Oh}, {Parejko}, {Parley}, {Pascual},
  {Patil}, {Patil}, {Plunkett}, {Prochaska}, {Rastogi}, {Reddy Janga},
  {Sabater}, {Sakurikar}, {Seifert}, {Sherbert}, {Sherwood-Taylor}, {Shih},
  {Sick}, {Silbiger}, {Singanamalla}, {Singer}, {Sladen}, {Sooley},
  {Sornarajah}, {Streicher}, {Teuben}, {Thomas}, {Tremblay}, {Turner},
  {Terr{\'o}n}, {van Kerkwijk}, {de la Vega}, {Watkins}, {Weaver}, {Whitmore},
  {Woillez}, {Zabalza}, \& {Astropy Contributors}}]{astropy2018}
{Astropy Collaboration}, {Price-Whelan}, A.~M., {Sip{\H{o}}cz}, B.~M., {et~al.}
  2018, \aj, 156, 123

\bibitem[{{Babul} \& {Rees}(1992)}]{Babul1992}
{Babul}, A., \& {Rees}, M.~J. 1992, \mnras, 255, 346

\bibitem[{{Bechtol} {et~al.}(2015){Bechtol}, {Drlica-Wagner}, {Balbinot},
  {Pieres}, {Simon}, {Yanny}, {Santiago}, {Wechsler}, {Frieman}, {Walker},
  {Williams}, {Rozo}, {Rykoff}, {Queiroz}, {Luque}, {Benoit-L{\'e}vy},
  {Tucker}, {Sevilla}, {Gruendl}, {da Costa}, {Fausti Neto}, {Maia}, {Abbott},
  {Allam}, {Armstrong}, {Bauer}, {Bernstein}, {Bernstein}, {Bertin}, {Brooks},
  {Buckley-Geer}, {Burke}, {Carnero Rosell}, {Castander}, {Covarrubias},
  {D'Andrea}, {DePoy}, {Desai}, {Diehl}, {Eifler}, {Estrada}, {Evrard},
  {Fernandez}, {Finley}, {Flaugher}, {Gaztanaga}, {Gerdes}, {Girardi},
  {Gladders}, {Gruen}, {Gutierrez}, {Hao}, {Honscheid}, {Jain}, {James},
  {Kent}, {Kron}, {Kuehn}, {Kuropatkin}, {Lahav}, {Li}, {Lin}, {Makler},
  {March}, {Marshall}, {Martini}, {Merritt}, {Miller}, {Miquel}, {Mohr},
  {Neilsen}, {Nichol}, {Nord}, {Ogando}, {Peoples}, {Petravick}, {Plazas},
  {Romer}, {Roodman}, {Sako}, {Sanchez}, {Scarpine}, {Schubnell}, {Smith},
  {Soares-Santos}, {Sobreira}, {Suchyta}, {Swanson}, {Tarle}, {Thaler},
  {Thomas}, {Wester}, {Zuntz}, \& {DES Collaboration}}]{Bechtol2015}
{Bechtol}, K., {Drlica-Wagner}, A., {Balbinot}, E., {et~al.} 2015, \apj, 807,
  50

\bibitem[{{Bedin} {et~al.}(2005){Bedin}, {Cassisi}, {Castelli}, {Piotto},
  {Anderson}, {Salaris}, {Momany}, \& {Pietrinferni}}]{Bedin2005}
{Bedin}, L.~R., {Cassisi}, S., {Castelli}, F., {et~al.} 2005, \mnras, 357, 1038

\bibitem[{{Belokurov} {et~al.}(2006){Belokurov}, {Zucker}, {Evans},
  {Wilkinson}, {Irwin}, {Hodgkin}, {Bramich}, {Irwin}, {Gilmore}, {Willman},
  {Vidrih}, {Newberg}, {Wyse}, {Fellhauer}, {Hewett}, {Cole}, {Bell}, {Beers},
  {Rockosi}, {Yanny}, {Grebel}, {Schneider}, {Lupton}, {Barentine},
  {Brewington}, {Brinkmann}, {Harvanek}, {Kleinman}, {Krzesinski}, {Long},
  {Nitta}, {Smith}, \& {Snedden}}]{Belokurov2006}
{Belokurov}, V., {Zucker}, D.~B., {Evans}, N.~W., {et~al.} 2006, \apjl, 647,
  L111

\bibitem[{{Ben{\'{\i}}tez-Llambay} {et~al.}(2015){Ben{\'{\i}}tez-Llambay},
  {Navarro}, {Abadi}, {Gottl{\"o}ber}, {Yepes}, {Hoffman}, \&
  {Steinmetz}}]{BenitezLlambay2015}
{Ben{\'{\i}}tez-Llambay}, A., {Navarro}, J.~F., {Abadi}, M.~G., {et~al.} 2015,
  \mnras, 450, 4207

\bibitem[{{Bermejo-Climent} {et~al.}(2018){Bermejo-Climent}, {Battaglia},
  {Gallart}, {Di Cintio}, {Brook}, {Cicu{\'e}ndez}, {Monelli}, {Leaman},
  {Mayer}, {Pe{\~n}arrubia}, \& {Read}}]{Bermejo2018}
{Bermejo-Climent}, J.~R., {Battaglia}, G., {Gallart}, C., {et~al.} 2018,
  \mnras, 479, 1514

\bibitem[{{Bernard} {et~al.}(2015){Bernard}, {Ferguson}, {Chapman}, {Ibata},
  {Irwin}, {Lewis}, \& {McConnachie}}]{Bernard2015letter}
{Bernard}, E.~J., {Ferguson}, A.~M.~N., {Chapman}, S.~C., {et~al.} 2015,
  \mnras, 453, L113

\bibitem[{{Bernard} {et~al.}(2018){Bernard}, {Schultheis}, {Di Matteo}, {Hill},
  {Haywood}, \& {Calamida}}]{Bernard2018}
{Bernard}, E.~J., {Schultheis}, M., {Di Matteo}, P., {et~al.} 2018, \mnras,
  477, 3507

\bibitem[{{Bland-Hawthorn} {et~al.}(2015){Bland-Hawthorn}, {Sutherland}, \&
  {Webster}}]{BH2015}
{Bland-Hawthorn}, J., {Sutherland}, R., \& {Webster}, D. 2015, \apj, 807, 154

\bibitem[{{Bovill} \& {Ricotti}(2009)}]{BovillRicotti2009}
{Bovill}, M.~S., \& {Ricotti}, M. 2009, \apj, 693, 1859

\bibitem[{{Bressan} {et~al.}(2012){Bressan}, {Marigo}, {Girardi}, {Salasnich},
  {Dal Cero}, {Rubele}, \& {Nanni}}]{Bressan2012}
{Bressan}, A., {Marigo}, P., {Girardi}, L., {et~al.} 2012, \mnras, 427, 127

\bibitem[{{Bromm} \& {Yoshida}(2011)}]{Bromm2011}
{Bromm}, V., \& {Yoshida}, N. 2011, \araa, 49, 373

\bibitem[{Brook {et~al.}(2014)Brook, Cintio, Knebe, Gottlöber, Hoffman, Yepes,
  \& Garrison-Kimmel}]{Brook2014}
Brook, C.~B., Cintio, A.~D., Knebe, A., {et~al.} 2014, The Astrophysical
  Journal, 784, L14

\bibitem[{{Brown} {et~al.}(2012){Brown}, {Tumlinson}, {Geha}, {Kirby},
  {VandenBerg}, {Mu{\~n}oz}, {Kalirai}, {Simon}, {Avila}, {Guhathakurta},
  {Renzini}, \& {Ferguson}}]{Brown2012}
{Brown}, T.~M., {Tumlinson}, J., {Geha}, M., {et~al.} 2012, \apjl, 753, L21

\bibitem[{{Brown} {et~al.}(2014){Brown}, {Tumlinson}, {Geha}, {Simon},
  {Vargas}, {VandenBerg}, {Kirby}, {Kalirai}, {Avila}, {Gennaro}, {Ferguson},
  {Mu{\~n}oz}, {Guhathakurta}, \& {Renzini}}]{Brown2014}
---. 2014, \apj, 796, 91

\bibitem[{{Buck} {et~al.}(2019){Buck}, {Macci{\`o}}, {Dutton}, {Obreja}, \&
  {Frings}}]{Buck2019}
{Buck}, T., {Macci{\`o}}, A.~V., {Dutton}, A.~A., {Obreja}, A., \& {Frings}, J.
  2019, \mnras, 483, 1314

\bibitem[{{Bullock} \& {Boylan-Kolchin}(2017)}]{Bullock2017ARAA}
{Bullock}, J.~S., \& {Boylan-Kolchin}, M. 2017, \araa, 55, 343

\bibitem[{{Cashmore} {et~al.}(2017){Cashmore}, {Wilkinson}, {Power}, \&
  {Bourne}}]{Cashmore2017}
{Cashmore}, C.~R., {Wilkinson}, M.~I., {Power}, C., \& {Bourne}, M. 2017,
  \mnras, 468, 451

\bibitem[{{Cassisi} \& {Salaris}(2013)}]{Cassisi_Salaris2013}
{Cassisi}, S., \& {Salaris}, M. 2013, {Old Stellar Populations: How to Study
  the Fossil Record of Galaxy Formation}

\bibitem[{{Chau} {et~al.}(2017){Chau}, {Mayer}, \& {Governato}}]{Chau2017}
{Chau}, A., {Mayer}, L., \& {Governato}, F. 2017, \apj, 845, 17

\bibitem[{{Cignoni} {et~al.}(2015){Cignoni}, {Sabbi}, {van der Marel}, {Tosi},
  {Zaritsky}, {Anderson}, {Lennon}, {Aloisi}, {de Marchi}, {Gouliermis},
  {Grebel}, {Smith}, \& {Zeidler}}]{Cignoni2015}
{Cignoni}, M., {Sabbi}, E., {van der Marel}, R.~P., {et~al.} 2015, \apj, 811,
  76

\bibitem[{{Clementini} {et~al.}(2012){Clementini}, {Cignoni}, {Contreras
  Ramos}, {Federici}, {Ripepi}, {Marconi}, {Tosi}, \&
  {Musella}}]{Clementini2012}
{Clementini}, G., {Cignoni}, M., {Contreras Ramos}, R., {et~al.} 2012, \apj,
  756, 108

\bibitem[{{Contenta} {et~al.}(2018){Contenta}, {Balbinot}, {Petts}, {Read},
  {Gieles}, {Collins}, {Pe{\~n}arrubia}, {Delorme}, \&
  {Gualandris}}]{Contenta2018Cluster}
{Contenta}, F., {Balbinot}, E., {Petts}, J.~A., {et~al.} 2018, \mnras, 476,
  3124

\bibitem[{{Cordier} {et~al.}(2007){Cordier}, {Pietrinferni}, {Cassisi}, \&
  {Salaris}}]{Cordier2007}
{Cordier}, D., {Pietrinferni}, A., {Cassisi}, S., \& {Salaris}, M. 2007, \aj,
  133, 468

\bibitem[{{Crnojevi{\'c}} {et~al.}(2016){Crnojevi{\'c}}, {Sand}, {Zaritsky},
  {Spekkens}, {Willman}, \& {Hargis}}]{Crnojevic2016}
{Crnojevi{\'c}}, D., {Sand}, D.~J., {Zaritsky}, D., {et~al.} 2016, \apjl, 824,
  L14

\bibitem[{{Di Cecco} {et~al.}(2010){Di Cecco}, {Becucci}, {Bono}, {Monelli},
  {Stetson}, {Degl'Innocenti}, {Prada Moroni}, {Nonino}, {Weiss}, {Buonanno},
  {Calamida}, {Caputo}, {Corsi}, {Ferraro}, {Iannicola}, {Pulone},
  {Romaniello}, \& {Walker}}]{DiCecco2010}
{Di Cecco}, A., {Becucci}, R., {Bono}, G., {et~al.} 2010, \pasp, 122, 991

\bibitem[{{Dolphin}(2002)}]{Dolphin2002}
{Dolphin}, A.~E. 2002, \mnras, 332, 91

\bibitem[{Fakhouri {et~al.}(2010)Fakhouri, Ma, \&
  Boylan-Kolchin}]{Fakhouri2010}
Fakhouri, O., Ma, C.-P., \& Boylan-Kolchin, M. 2010, Monthly Notices of the
  Royal Astronomical Society, 406, 2267–2278

\bibitem[{{Fitts} {et~al.}(2017){Fitts}, {Boylan-Kolchin}, {Elbert}, {Bullock},
  {Hopkins}, {O{\~n}orbe}, {Wetzel}, {Wheeler}, {Faucher-Gigu{\`e}re},
  {Kere{\v{s}}}, {Skillman}, \& {Weisz}}]{Fitts2017}
{Fitts}, A., {Boylan-Kolchin}, M., {Elbert}, O.~D., {et~al.} 2017, \mnras, 471,
  3547

\bibitem[{{Ford} {et~al.}(1998){Ford}, {Bartko}, {Bely}, {Broadhurst},
  {Burrows}, {Cheng}, {Clampin}, {Crocker}, {Feldman}, {Golimowski}, {Hartig},
  {Illingworth}, {Kimble}, {Lesser}, {Miley}, {Neff}, {Postman}, {Sparks},
  {Tsvetanov}, {White}, {Sullivan}, {Krebs}, {Leviton}, {La Jeunesse},
  {Burmester}, {Fike}, {Johnson}, {Slusher}, {Volmer}, \&
  {Woodruff}}]{Ford1998}
{Ford}, H.~C., {Bartko}, F., {Bely}, P.~Y., {et~al.} 1998, in \procspie, Vol.
  3356, Space Telescopes and Instruments V, ed. P.~Y. {Bely} \& J.~B.
  {Breckinridge}, 234--248

\bibitem[{{Fritz} {et~al.}(2018){Fritz}, {Battaglia}, {Pawlowski},
  {Kallivayalil}, {van der Marel}, {Sohn}, {Brook}, \& {Besla}}]{Fritz2018}
{Fritz}, T.~K., {Battaglia}, G., {Pawlowski}, M.~S., {et~al.} 2018, \aap, 619,
  A103

\bibitem[{{Gallart} {et~al.}(1996){Gallart}, {Aparicio}, \&
  {Vilchez}}]{Gallart1996a}
{Gallart}, C., {Aparicio}, A., \& {Vilchez}, J.~M. 1996, \aj, 112, 1928

\bibitem[{{Gallart} {et~al.}(2005){Gallart}, {Zoccali}, \&
  {Aparicio}}]{Gallart2005ARAA}
{Gallart}, C., {Zoccali}, M., \& {Aparicio}, A. 2005, \araa, 43, 387

\bibitem[{{Gallart} {et~al.}(1999){Gallart}, {Freedman}, {Mateo}, {Chiosi},
  {Thompson}, {Aparicio}, {Bertelli}, {Hodge}, {Lee}, {Olszewski}, {Saha},
  {Stetson}, \& {Suntzeff}}]{Gallart1999data}
{Gallart}, C., {Freedman}, W.~L., {Mateo}, M., {et~al.} 1999, \apj, 514, 665

\bibitem[{{Gallart} {et~al.}(2015){Gallart}, {Monelli}, {Mayer}, {Aparicio},
  {Battaglia}, {Bernard}, {Cassisi}, {Cole}, {Dolphin}, {Drozdovsky},
  {Hidalgo}, {Navarro}, {Salvadori}, {Skillman}, {Stetson}, \&
  {Weisz}}]{Gallart2015}
{Gallart}, C., {Monelli}, M., {Mayer}, L., {et~al.} 2015, \apjl, 811, L18

\bibitem[{{Gelli} {et~al.}(2020){Gelli}, {Salvadori}, {Pallottini}, \&
  {Ferrara}}]{Gelli2020}
{Gelli}, V., {Salvadori}, S., {Pallottini}, A., \& {Ferrara}, A. 2020, \mnras,
  498, 4134

\bibitem[{Harris {et~al.}(2020)Harris, Millman, van~der Walt, Gommers,
  Virtanen, Cournapeau, Wieser, Taylor, Berg, Smith, {et~al.}}]{numpy2020}
Harris, C.~R., Millman, K.~J., van~der Walt, S.~J., {et~al.} 2020, Nature, 585,
  357

\bibitem[{{Helmi}(2020)}]{Helmi2020ARAA}
{Helmi}, A. 2020, \araa, 58, 205

\bibitem[{{Hidalgo} {et~al.}(2011){Hidalgo}, {Aparicio}, {Skillman}, {Monelli},
  {Gallart}, {Cole}, {Dolphin}, {Weisz}, {Bernard}, {Cassisi}, {Mayer},
  {Stetson}, {Tolstoy}, \& {Ferguson}}]{Hidalgo2011sfhlgs3}
{Hidalgo}, S.~L., {Aparicio}, A., {Skillman}, E., {et~al.} 2011, \apj, 730, 14

\bibitem[{{Hidalgo} {et~al.}(2013){Hidalgo}, {Monelli}, {Aparicio}, {Gallart},
  {Skillman}, {Cassisi}, {Bernard}, {Mayer}, {Stetson}, {Cole}, \&
  {Dolphin}}]{Hidalgo2013}
{Hidalgo}, S.~L., {Monelli}, M., {Aparicio}, A., {et~al.} 2013, \apj, 778, 103

\bibitem[{Hunter(2007)}]{Hunter2007}
Hunter, J.~D. 2007, Computing in Science \& Engineering, 9, 90

\bibitem[{{Irwin} {et~al.}(2007){Irwin}, {Belokurov}, {Evans}, {Ryan-Weber},
  {de Jong}, {Koposov}, {Zucker}, {Hodgkin}, {Gilmore}, {Prema}, {Hebb},
  {Begum}, {Fellhauer}, {Hewett}, {Kennicutt}, {Wilkinson}, {Bramich},
  {Vidrih}, {Rix}, {Beers}, {Barentine}, {Brewington}, {Harvanek},
  {Krzesinski}, {Long}, {Nitta}, \& {Snedden}}]{Irwin2007LeoT}
{Irwin}, M.~J., {Belokurov}, V., {Evans}, N.~W., {et~al.} 2007, \apjl, 656, L13

\bibitem[{{Janesh} {et~al.}(2019){Janesh}, {Rhode}, {Salzer}, {Janowiecki},
  {Adams}, {Haynes}, {Giovanelli}, \& {Cannon}}]{Janesh2019_HI_UFD}
{Janesh}, W., {Rhode}, K.~L., {Salzer}, J.~J., {et~al.} 2019, \aj, 157, 183

\bibitem[{{Jeon} {et~al.}(2017){Jeon}, {Besla}, \& {Bromm}}]{Jeon2017}
{Jeon}, M., {Besla}, G., \& {Bromm}, V. 2017, \apj, 848, 85

\bibitem[{{Jeon} {et~al.}(2015){Jeon}, {Bromm}, {Pawlik}, \&
  {Milosavljevi{\'c}}}]{Jeon2015}
{Jeon}, M., {Bromm}, V., {Pawlik}, A.~H., \& {Milosavljevi{\'c}}, M. 2015,
  \mnras, 452, 1152

\bibitem[{{Katz} {et~al.}(2020){Katz}, {Ramsoy}, {Rosdahl}, {Kimm}, {Blaizot},
  {Haehnelt}, {Michel-Dansac}, {Garel}, {Laigle}, {Devriendt}, \&
  {Slyz}}]{Katz2020}
{Katz}, H., {Ramsoy}, M., {Rosdahl}, J., {et~al.} 2020, \mnras, 494, 2200

\bibitem[{{Koposov} {et~al.}(2015){Koposov}, {Belokurov}, {Torrealba}, \&
  {Evans}}]{Koposov2015a}
{Koposov}, S.~E., {Belokurov}, V., {Torrealba}, G., \& {Evans}, N.~W. 2015,
  \apj, 805, 130

\bibitem[{{Kravtsov} {et~al.}(2004){Kravtsov}, {Gnedin}, \&
  {Klypin}}]{Kravtsov2004}
{Kravtsov}, A.~V., {Gnedin}, O.~Y., \& {Klypin}, A.~A. 2004, \apj, 609, 482

\bibitem[{{Kroupa}(2001)}]{Kroupa2001}
{Kroupa}, P. 2001, \mnras, 322, 231

\bibitem[{{Kroupa} {et~al.}(1993){Kroupa}, {Tout}, \& {Gilmore}}]{Kroupa1993}
{Kroupa}, P., {Tout}, C.~A., \& {Gilmore}, G. 1993, \mnras, 262, 545

\bibitem[{{Li} {et~al.}(2017){Li}, {Simon}, {Drlica-Wagner}, {Bechtol}, {Wang},
  {Garc{\'{\i}}a-Bellido}, {Frieman}, {Marshall}, {James}, {Strigari}, {Pace},
  {Balbinot}, {Zhang}, {Abbott}, {Allam}, {Benoit-L{\'e}vy}, {Bernstein},
  {Bertin}, {Brooks}, {Burke}, {Carnero Rosell}, {Carrasco Kind}, {Carretero},
  {Cunha}, {D'Andrea}, {da Costa}, {DePoy}, {Desai}, {Diehl}, {Eifler},
  {Flaugher}, {Goldstein}, {Gruen}, {Gruendl}, {Gschwend}, {Gutierrez},
  {Krause}, {Kuehn}, {Lin}, {Maia}, {March}, {Menanteau}, {Miquel}, {Plazas},
  {Romer}, {Sanchez}, {Santiago}, {Schubnell}, {Sevilla-Noarbe}, {Smith},
  {Sobreira}, {Suchyta}, {Tarle}, {Thomas}, {Tucker}, {Walker}, {Wechsler},
  {Wester}, {Yanny}, \& {(DES Collaboration}}]{Li2017}
{Li}, T.~S., {Simon}, J.~D., {Drlica-Wagner}, A., {et~al.} 2017, \apj, 838, 8

\bibitem[{{Mac Low} \& {Ferrara}(1999)}]{MacLowFerrara1999}
{Mac Low}, M.-M., \& {Ferrara}, A. 1999, \apj, 513, 142

\bibitem[{{Marigo} {et~al.}(2017){Marigo}, {Girardi}, {Bressan}, {Rosenfield},
  {Aringer}, {Chen}, {Dussin}, {Nanni}, {Pastorelli}, {Rodrigues}, {Trabucchi},
  {Bladh}, {Dalcanton}, {Groenewegen}, {Montalb{\'a}n}, \& {Wood}}]{Marigo2017}
{Marigo}, P., {Girardi}, L., {Bressan}, A., {et~al.} 2017, \apj, 835, 77

\bibitem[{{Marsh} \& {Niemeyer}(2019)}]{Marsh2019Cluster}
{Marsh}, D. J.~E., \& {Niemeyer}, J.~C. 2019, \prl, 123, 051103

\bibitem[{{Martin} {et~al.}(2015){Martin}, {Nidever}, {Besla}, {Olsen},
  {Walker}, {Vivas}, {Gruendl}, {Kaleida}, {Mu{\~n}oz}, {Blum}, {Saha}, {Conn},
  {Bell}, {Chu}, {Cioni}, {de Boer}, {Gallart}, {Jin}, {Kunder}, {Majewski},
  {Martinez-Delgado}, {Monachesi}, {Monelli}, {Monteagudo}, {No{\"e}l},
  {Olszewski}, {Stringfellow}, {van der Marel}, \&
  {Zaritsky}}]{Martin2015Hydra}
{Martin}, N.~F., {Nidever}, D.~L., {Besla}, G., {et~al.} 2015, \apjl, 804, L5

\bibitem[{{Mayer}(2010)}]{Mayer2010}
{Mayer}, L. 2010, Advances in Astronomy, 2010, arXiv:0909.4075

\bibitem[{{McQuinn} {et~al.}(2015){McQuinn}, {Skillman}, {Dolphin}, {Cannon},
  {Salzer}, {Rhode}, {Adams}, {Berg}, {Giovanelli}, {Girardi}, \&
  {Haynes}}]{McQuinn2015LeoP_ACS}
{McQuinn}, K. B.~W., {Skillman}, E.~D., {Dolphin}, A., {et~al.} 2015, \apj,
  812, 158

\bibitem[{{Mina} {et~al.}(2020){Mina}, {Shen}, {Keller}, {Mayer}, {Madau}, \&
  {Wadsley}}]{Mina2020arXiv}
{Mina}, M., {Shen}, S., {Keller}, B.~W., {et~al.} 2020, arXiv e-prints,
  arXiv:2009.06646

\bibitem[{{Mocz} {et~al.}(2020){Mocz}, {Fialkov}, {Vogelsberger}, {Becerra},
  {Shen}, {Robles}, {Amin}, {Zavala}, {Boylan-Kolchin}, {Bose}, {Marinacci},
  {Chavanis}, {Lancaster}, \& {Hernquist}}]{Mocz2020}
{Mocz}, P., {Fialkov}, A., {Vogelsberger}, M., {et~al.} 2020, \mnras, 494, 2027

\bibitem[{{Monelli} {et~al.}(2010{\natexlab{a}}){Monelli}, {Hidalgo},
  {Stetson}, {Aparicio}, {Gallart}, {Dolphin}, {Cole}, {Weisz}, {Skillman},
  {Bernard}, {Mayer}, {Navarro}, {Cassisi}, {Drozdovsky}, \&
  {Tolstoy}}]{Monelli2010sfhcetus}
{Monelli}, M., {Hidalgo}, S.~L., {Stetson}, P.~B., {et~al.} 2010{\natexlab{a}},
  \apj, 720, 1225

\bibitem[{{Monelli} {et~al.}(2010{\natexlab{b}}){Monelli}, {Gallart},
  {Hidalgo}, {Aparicio}, {Skillman}, {Cole}, {Weisz}, {Mayer}, {Bernard},
  {Cassisi}, {Dolphin}, {Drozdovsky}, \& {Stetson}}]{Monelli2010sfhtucana}
{Monelli}, M., {Gallart}, C., {Hidalgo}, S.~L., {et~al.} 2010{\natexlab{b}},
  \apj, 722, 1864

\bibitem[{{Monelli} {et~al.}(2012){Monelli}, {Cassisi}, {Mapelli}, {Bernard},
  {Aparicio}, {Skillman}, {Stetson}, {Gallart}, {Hidalgo}, {Mayer}, \&
  {Tolstoy}}]{Monelli2012bss}
{Monelli}, M., {Cassisi}, S., {Mapelli}, M., {et~al.} 2012, \apj, 744, 157

\bibitem[{{Moster} {et~al.}(2013){Moster}, {Naab}, \& {White}}]{Moster2013}
{Moster}, B.~P., {Naab}, T., \& {White}, S. D.~M. 2013, \mnras, 428, 3121

\bibitem[{{Pallottini} {et~al.}(2017){Pallottini}, {Ferrara}, {Bovino},
  {Vallini}, {Gallerani}, {Maiolino}, \& {Salvadori}}]{Pallottini2017}
{Pallottini}, A., {Ferrara}, A., {Bovino}, S., {et~al.} 2017, \mnras, 471, 4128

\bibitem[{{Pietrinferni} {et~al.}(2004){Pietrinferni}, {Cassisi}, {Salaris}, \&
  {Castelli}}]{Pietrinferni2004}
{Pietrinferni}, A., {Cassisi}, S., {Salaris}, M., \& {Castelli}, F. 2004, \apj,
  612, 168

\bibitem[{{Pietrinferni} {et~al.}(2006){Pietrinferni}, {Cassisi}, {Salaris}, \&
  {Castelli}}]{Pietrinferni2006}
---. 2006, \apj, 642, 797

\bibitem[{{Revaz} \& {Jablonka}(2018)}]{RevazJablonka2018}
{Revaz}, Y., \& {Jablonka}, P. 2018, \aap, 616, A96

\bibitem[{{Rey} {et~al.}(2020){Rey}, {Pontzen}, {Agertz}, {Orkney}, {Read}, \&
  {Rosdahl}}]{Rey2020}
{Rey}, M.~P., {Pontzen}, A., {Agertz}, O., {et~al.} 2020, \mnras, 497, 1508

\bibitem[{{Ricotti}(2009)}]{Ricotti2009}
{Ricotti}, M. 2009, \mnras, 392, L45

\bibitem[{{Ricotti} \& {Gnedin}(2005)}]{Ricotti2005}
{Ricotti}, M., \& {Gnedin}, N.~Y. 2005, \apj, 629, 259

\bibitem[{{Ricotti} {et~al.}(2002){Ricotti}, {Gnedin}, \&
  {Shull}}]{Ricotti2002}
{Ricotti}, M., {Gnedin}, N.~Y., \& {Shull}, J.~M. 2002, \apj, 575, 49

\bibitem[{{Romano} {et~al.}(2019){Romano}, {Calura}, {D'Ercole}, \&
  {Few}}]{Romano2019}
{Romano}, D., {Calura}, F., {D'Ercole}, A., \& {Few}, C.~G. 2019, \aap, 630,
  A140

\bibitem[{{Ruiz-Lara} {et~al.}(2020){Ruiz-Lara}, {Gallart}, {Monelli}, {Fritz},
  {Battaglia}, {Cassisi}, {Aznar}, {Russo Cabrera},
  {Rodr{\'\i}guez-Mart{\'\i}n}, \&
  {Salazar-Gonz{\'a}lez}}]{Ruiz-Lara2020MNRAStmp}
{Ruiz-Lara}, T., {Gallart}, C., {Monelli}, M., {et~al.} 2020, \mnras,
  arXiv:2012.07863

\bibitem[{{Ryan-Weber} {et~al.}(2008){Ryan-Weber}, {Begum}, {Oosterloo}, {Pal},
  {Irwin}, {Belokurov}, {Evans}, \& {Zucker}}]{RyanWeber2008}
{Ryan-Weber}, E.~V., {Begum}, A., {Oosterloo}, T., {et~al.} 2008, \mnras, 384,
  535

\bibitem[{{Salvadori} \& {Ferrara}(2009)}]{SalvadoriFerrara2009}
{Salvadori}, S., \& {Ferrara}, A. 2009, \mnras, 395, L6

\bibitem[{{Salvadori} {et~al.}(2008){Salvadori}, {Ferrara}, \&
  {Schneider}}]{Salvadori2008}
{Salvadori}, S., {Ferrara}, A., \& {Schneider}, R. 2008, \mnras, 386, 348

\bibitem[{{Salvadori} {et~al.}(2015){Salvadori}, {Sk{\'u}lad{\'o}ttir}, \&
  {Tolstoy}}]{Salvadori2015}
{Salvadori}, S., {Sk{\'u}lad{\'o}ttir}, {\'A}., \& {Tolstoy}, E. 2015, \mnras,
  454, 1320

\bibitem[{{Salvadori} {et~al.}(2014){Salvadori}, {Tolstoy}, {Ferrara}, \&
  {Zaroubi}}]{Salvadori2014}
{Salvadori}, S., {Tolstoy}, E., {Ferrara}, A., \& {Zaroubi}, S. 2014, \mnras,
  437, L26

\bibitem[{{Sawala} {et~al.}(2010){Sawala}, {Scannapieco}, {Maio}, \&
  {White}}]{Sawala2010}
{Sawala}, T., {Scannapieco}, C., {Maio}, U., \& {White}, S. 2010, \mnras, 402,
  1599

\bibitem[{{Schlafly} \& {Finkbeiner}(2011)}]{Schlafly2011}
{Schlafly}, E.~F., \& {Finkbeiner}, D.~P. 2011, \apj, 737, 103

\bibitem[{{Shen} {et~al.}(2014){Shen}, {Madau}, {Conroy}, {Governato}, \&
  {Mayer}}]{Shen2014}
{Shen}, S., {Madau}, P., {Conroy}, C., {Governato}, F., \& {Mayer}, L. 2014,
  \apj, 792, 99

\bibitem[{{Simon} \& {Geha}(2007)}]{SimonGeha2007}
{Simon}, J.~D., \& {Geha}, M. 2007, \apj, 670, 313

\bibitem[{{Simon} {et~al.}(2020){Simon}, {Brown}, {Drlica-Wagner}, {Li},
  {Avila}, {Bechtol}, {Clementini}, {Crnojevic}, {Garofalo}, {Geha}, {Sand},
  {Strader}, \& {Willman}}]{Simon2020arXiv}
{Simon}, J.~D., {Brown}, T.~M., {Drlica-Wagner}, A., {et~al.} 2020, arXiv
  e-prints, arXiv:2012.00043

\bibitem[{{Stetson}(1987)}]{Stetson87}
{Stetson}, P.~B. 1987, \pasp, 99, 191

\bibitem[{{Stetson}(1993)}]{Stetson1993ASP}
{Stetson}, P.~B. 1993, in Astronomical Society of the Pacific Conference
  Series, Vol.~48, The Globular Cluster-Galaxy Connection, ed. G.~H. {Smith} \&
  J.~P. {Brodie}, 14

\bibitem[{{Stetson}(1994)}]{Stetson94}
---. 1994, \pasp, 106, 250

\bibitem[{{Stinson} {et~al.}(2009){Stinson}, {Dalcanton}, {Quinn}, {Gogarten},
  {Kaufmann}, \& {Wadsley}}]{Stinson2009}
{Stinson}, G.~S., {Dalcanton}, J.~J., {Quinn}, T., {et~al.} 2009, \mnras, 395,
  1455

\bibitem[{{Tang} {et~al.}(2014){Tang}, {Bressan}, {Rosenfield}, {Slemer},
  {Marigo}, {Girardi}, \& {Bianchi}}]{Tang2014}
{Tang}, J., {Bressan}, A., {Rosenfield}, P., {et~al.} 2014, \mnras, 445, 4287

\bibitem[{{VandenBerg} {et~al.}(2016){VandenBerg}, {Denissenkov}, \&
  {Catelan}}]{VandenBerg2016}
{VandenBerg}, D.~A., {Denissenkov}, P.~A., \& {Catelan}, M. 2016, \apj, 827, 2

\bibitem[{Virtanen {et~al.}(2020)Virtanen, Gommers, Oliphant, Haberland, Reddy,
  Cournapeau, Burovski, Peterson, Weckesser, Bright, {van der Walt}, Brett,
  Wilson, Millman, Mayorov, Nelson, Jones, Kern, Larson, Carey, Polat, Feng,
  Moore, {VanderPlas}, Laxalde, Perktold, Cimrman, Henriksen, Quintero, Harris,
  Archibald, Ribeiro, Pedregosa, {van Mulbregt}, \& {SciPy 1.0
  Contributors}}]{scipy2020}
Virtanen, P., Gommers, R., Oliphant, T.~E., {et~al.} 2020, Nature Methods, 17,
  261

\bibitem[{{Weisz} {et~al.}(2012){Weisz}, {Zucker}, {Dolphin}, {Martin}, {de
  Jong}, {Holtzman}, {Dalcanton}, {Gilbert}, {Williams}, {Bell}, {Belokurov},
  \& {Wyn Evans}}]{Weisz2012LeoT}
{Weisz}, D.~R., {Zucker}, D.~B., {Dolphin}, A.~E., {et~al.} 2012, \apj, 748, 88

\bibitem[{{Westmeier} {et~al.}(2015){Westmeier}, {Staveley-Smith},
  {Calabretta}, {Jurek}, {Koribalski}, {Meyer}, {Popping}, \&
  {Wong}}]{Westmeier2015}
{Westmeier}, T., {Staveley-Smith}, L., {Calabretta}, M., {et~al.} 2015, \mnras,
  453, 338

\bibitem[{{Wheeler} {et~al.}(2015){Wheeler}, {O{\~n}orbe}, {Bullock},
  {Boylan-Kolchin}, {Elbert}, {Garrison-Kimmel}, {Hopkins}, \&
  {Kere{\v{s}}}}]{Wheeler2015}
{Wheeler}, C., {O{\~n}orbe}, J., {Bullock}, J.~S., {et~al.} 2015, \mnras, 453,
  1305

\bibitem[{{Wheeler} {et~al.}(2019){Wheeler}, {Hopkins}, {Pace},
  {Garrison-Kimmel}, {Boylan-Kolchin}, {Wetzel}, {Bullock}, {Kere{\v{s}}},
  {Faucher-Gigu{\`e}re}, \& {Quataert}}]{Wheeler2019}
{Wheeler}, C., {Hopkins}, P.~F., {Pace}, A.~B., {et~al.} 2019, \mnras, 490,
  4447

\bibitem[{{Willman} {et~al.}(2005){Willman}, {Dalcanton}, {Martinez-Delgado},
  {West}, {Blanton}, {Hogg}, {Barentine}, {Brewington}, {Harvanek}, {Kleinman},
  {Krzesinski}, {Long}, {Neilsen}, {Nitta}, \& {Snedden}}]{Willman2005b}
{Willman}, B., {Dalcanton}, J.~J., {Martinez-Delgado}, D., {et~al.} 2005,
  \apjl, 626, L85

\bibitem[{Woo {et~al.}(2008)Woo, Courteau, \& Dekel}]{Woo2008}
Woo, J., Courteau, S., \& Dekel, A. 2008, Monthly Notices of the Royal
  Astronomical Society, doi:10.1111/j.1365-2966.2008.13770.x

\bibitem[{{Woo} {et~al.}(2003){Woo}, {Gallart}, {Demarque}, {Yi}, \&
  {Zoccali}}]{Woo2003}
{Woo}, J.-H., {Gallart}, C., {Demarque}, P., {Yi}, S., \& {Zoccali}, M. 2003,
  \aj, 125, 754

\bibitem[{{Zoutendijk} {et~al.}(2020){Zoutendijk}, {Brinchmann}, {Boogaard},
  {Gunawardhana}, {Husser}, {Kamann}, {Ramos Padilla}, {Roth}, {Bacon}, {den
  Brok}, {Dreizler}, \& {Krajnovi{\'c}}}]{Zoutendijk2020cluster}
{Zoutendijk}, S.~L., {Brinchmann}, J., {Boogaard}, L.~A., {et~al.} 2020, \aap,
  635, A107

\bibitem[{{Zucker} {et~al.}(2006){Zucker}, {Belokurov}, {Evans}, {Kleyna},
  {Irwin}, {Wilkinson}, {Fellhauer}, {Bramich}, {Gilmore}, {Newberg}, {Yanny},
  {Smith}, {Hewett}, {Bell}, {Rix}, {Gnedin}, {Vidrih}, {Wyse}, {Willman},
  {Grebel}, {Schneider}, {Beers}, {Kniazev}, {Barentine}, {Brewington},
  {Brinkmann}, {Harvanek}, {Kleinman}, {Krzesinski}, {Long}, {Nitta}, \&
  {Snedden}}]{Zucker2006a}
{Zucker}, D.~B., {Belokurov}, V., {Evans}, N.~W., {et~al.} 2006, \apjl, 650,
  L41

\end{thebibliography}



\end{document}